\documentclass[preprint,eqsecnum,aps,keywords,nofootinbib]{revtex4}
\usepackage{graphicx,amssymb,amsmath}
\usepackage{epsfig}
\usepackage{float}
\usepackage{pdfpages}
\usepackage{latexsym}
\usepackage{rotating}
\usepackage{enumitem}
\begin{document}
\title[Shorttitle]{\large Null Geodesics and Red-Blue Shifts of Photons Emitted from Geodesic Particles Around a Non-Commutative Black Hole Spacetime}
\author{Ravi Shankar Kuniyal $^{a}$} \email{ravikuniyal09@gmail.com}
\author{Rashmi Uniyal $^{b}$} \email{rashmiuniyal001@gmail.com}
\author{Anindya Biswas $^{c}$} \email{ani_imsc@yahoo.co.in}
\author{Hemwati Nandan $^{d}$}\email{hnandan@associates.iucaa.in}
\author{K. D. Purohit $^{e}$} \email{kdpurohit@rediffmail.com}
\affiliation{$^{a, d}$ Department of Physics, Gurukula Kangri Vishwavidyalaya, Haridwar 249 404, Uttarakhand, India}
\affiliation{$^{b}$ Department of Physics, Govt. Degree College Narendranagar,
Tehri Garhwal - 249 175, Uttarakhand, India}
\affiliation{$^{c}$ Department of Physics, Ranaghat College, Ranaghat, West Bengal 741 201, India}
\affiliation{$^{e}$ Department of Physics, HNB Garhwal University, Srinagar, Garhwal 246 174, Uttarakhand, India}
\date{\today}
\begin{abstract}
\noindent We investigate the geodesic motion of massless test particles in the background of a non-commutative geometry inspired Schwarzschild black hole. The behaviour of effective potential is analysed in the equatorial plane and the possible motions of massless particles (i.e. photons) for different values of impact parameter are discussed accordingly. We have also calculated the frequency shift of photons in this spacetime. Further, the mass parameter of a non-commutative inspired Schwarzschild black hole is computed in terms of the measurable redshift of photons emitted by massive particles moving along circular geodesics in equatorial plane. It is observed that the the gravitational field of a non-commutative inspired Schwarzschild black hole is more attractive than the Schwarzschild black hole in General Relativity.
\end{abstract}
\maketitle
\section{Introduction}
\noindent In General Relativity (GR) black holes (BHs) are obtained as exact solutions of Einstein's field equations \cite{har, wal, psj, cha, epo}. In GR, the curvature singularities which are best exemplified by the BHs emerge in a way such that the classical description of the gravitational field breaks down. In order to obtain physically reliable scenarios in spacetime regions teased by such curvature singularities, one has the only possibility of rendering some formulation of the quantum theory of gravitation. The noncommutativity is a mathematical framework which provides us insight into a theory of gravity which is compatible with quantum mechanics by successfully removing the singularities those appear in GR \cite{nic}. This is possible by replacing the position Dirac-delta function with a Gaussian distribution
that presents a standard minimal width. In this way, the divergences that appear in GR can be avoided because noncommutativity will replace point like structures by smeared objects \cite{rah, piy}.\\
The study of geodesic structure of massless particles in a given black hole (BH) spacetime is one
of the important ways to understand the gravitational field around a BH spactime. The geodesic motion around various spacetimes in diverse contexts (for timelike as well as null geodesics), both in GR and  alternative theories of gravity, are studied widely time and again \cite{RefP1, RefF1, RefN, RefD1, RefF2, RefF3, RefK1, RefB, RefP2, RefP3, RefZ, RefJ, RefK2, RefU, RefU1, RefE1, RefT1, RefU2, RefRS, RefF4, RefE2, RefH1, RefA1, RefA3, RefG1, RefR, RefR11, ras22, ehm1, ehm, dab, hnd, val, sas1, sas2, rpk, raji, kaif, sahe}.\\
In the last few years, various
BH solutions inspired by a non-commutative geometry were found and important investigations have been done concerning these BH spacetimes \cite{arr, kli, bro, sma, gre}. The study of non-commutative BHs is of interest because they yields interesting information
about their behavior and properties and also aquires a deeper relevance in the research of the quantum nature of spacetime at very high energy scales. Ansoldi et al. derived a new non-commutative geometry inspired solution of the coupled Einstein Maxwell field equations describing a variety of charged, self-gravitating objects, including extremal and non-extremal BHs \cite{anso}.  Reissner Nordstr$\ddot{o}$m BH in non-commutative spaces is also studied by Alavi \cite{alav}. The tiny effect of noncommutativity on the efficiency of the Penrose process of rotational energy extraction from a BH is studied in the background of non-commutative Kerr BH by Grezia et al. \cite{grezi}. The behavior of the charged rotating non-commutative BHs was studied by Modesto et al. \cite{modesto}. Chabab et al. studied the Schwarzschild BHs in a D-dimensional non-commutative Space. In this work, various aspects related to the non-commutative extension are discussed and some non trivial results are derived \cite{chabab}. 
Recently, Rahaman et al. have constructed a BTZ BH solution in non-commutative geometry \cite{rah1}. Further, Rahaman et al. have also studied the motion of massive and massless test particles in the gravitational field of a non-commutative Schwarzschild BH \cite{rah}. The motion of the test particles in static equilibrium as well as non-static equilibrium are discussed using Hamilton-Jacobi formalism in \cite{rah}.
\\
This paper is organised as follows. In section II, a brief introduction to the spacetime considered is presented. In section III, we have discussed the geodesics of massless particles along with the nature of the effective potential followed by the structure of photon orbits. The study of the frequency shift of photons emitted by massive particles moving along circular geodesic in equatorial plane is presented in section IV. In the last section, we have summarized our results with future possibilities. It is observed that due to the presence of noncommutitavity, the gravitation attraction for NCBH spacetime is found weaker than that of a SBH spacetime in GR.
\section{Non-commutative black hole Spacetime}
\begin{figure}
\includegraphics[scale=1]{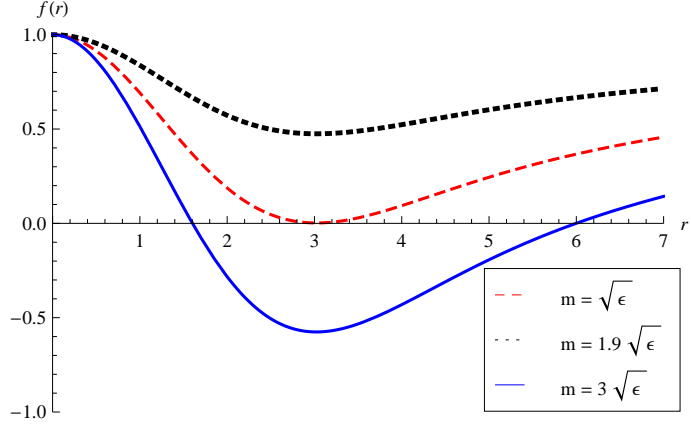}
\caption{Behaviour of $f(r)$ for various values of $\frac{m}{\sqrt{\epsilon}}$. There exist three different cases, namely two horizons (solid line),
no horizon (dotted line) and one single degenerate horizon (dashed line).}
\label{v1}
\end{figure}

\noindent A non-commutative Schwarzschild black hole (NCBH) spacetime \cite{nic} is described by
\begin{equation}
ds^2 = - f(r)  \hspace{0.1cm} dt^2 + \frac{dr^2}{f(r)} + r^2  \hspace{0.1cm}[ d\theta^2 + sin^2(\theta) \hspace{0.1cm}d\phi^2 ],
\label{BH1}
\end{equation}
where,
\begin{equation}
f(r) = 1 - \frac{4 \hspace{0.1cm} m}{r  \hspace{0.1cm}\sqrt{\pi}} \gamma\left(\frac{3}{2}, \frac{r^2}{4  \hspace{0.1cm}\epsilon}\right).
\label{gamma}
\end{equation}
Here, $\gamma$ is the lower incomplete gamma function that has the expression,
\begin{equation}
\gamma\left(\frac{3}{2}, \frac{r^2}{4 \epsilon}\right) \equiv \int_{0}^{\frac{r^2}{4 \epsilon}} \sqrt{t} \hspace{0.1cm} e^{-t}  \hspace{0.1cm}dt,
\end{equation}
and $\sqrt{\epsilon}$ is the width of the Gaussian mass-energy distribution of the source.
In the limit $\frac{r}{\sqrt{\epsilon}} \rightarrow \infty$, the NCBH metric reduces into a SBH spacetime. The horizon radius $(r_h)$ is given by $f(r_h) = 0$, which leads to,
\begin{equation}
 r_h = \frac{4 \hspace{0.1cm} m}{ \sqrt{\pi}} \gamma\left(\frac{3}{2}, \frac{r^2_{h}}{4  \hspace{0.1cm}\epsilon}\right).
\label{horizon}
\end{equation}
The Eqn.(\ref{horizon}) cannot be solved in a closed form. However, the existence and position of horizon can be visualised numerically from Fig.(\ref{v1}). Here, noncommutativity introduces a new behavior with respect to the standard SBH. The intercept on the horizontal axes give the radii of three different possible horizons \cite{nic} as listed below:\\
(i) One degenerate horizon for $ m = 1.9 \hspace{0.1cm} \sqrt{\epsilon}$ ;\\
(ii) Two distinct horizons (one inner horizon and one outer event horizon) for $ m = 3\hspace{0.1cm} \sqrt{\epsilon}$ ;\\
(iii) No horizon for $ m = \sqrt{\epsilon}$.\\
Further, in terms of upper incomplete gamma function, the Eqn.(\ref{horizon}) reads as,
\begin{equation}
r_h = 2 m \Big[ 1 - \frac{2}{\sqrt{\pi}}\hspace{0.1cm} \Gamma \left(\frac{3}{2}, \frac{m^2}{\epsilon}\right)\Big].
\end{equation}
Here, the first term corresponds to the Schwarzschild radius while second term is a correction term with $\epsilon$. At first order in $m/ \sqrt{\epsilon}$,
\begin{equation}
r_h = 2 \hspace{0.1cm} m  \hspace{0.1cm}\Big( 1 - \frac{m}{\sqrt{\pi \epsilon}} \hspace{0.1cm} e^{-\frac{m^2}{\epsilon}}\Big).
\end{equation}
One can observe that the effect of noncommutativity is very small at large distances, while at short distances, the quantum geometry becomes quite important.
\section{Null geodesics}
\noindent In this section, we study the radial and non-radial null geodesics alongwith the possible orbit structure for spacetime given by Eqn.(\ref{BH1}) in the equatorial plane (i.e. $\theta=\pi/2$).
The geodesic equations and constraint equation for null geodesics are given as, $\ddot{x^a}+\Gamma^{a}_{bc}\dot{x}^{b}\dot{x}^{c}=0 $ and  $g_{{a}{b}}\dot{x}^a \dot{x}^b=0$ respectively \cite{cha} (where dot denotes the differentiation with respect to an affine parameter).
Now using the geodesic equations, their first integrals can be obtained as,
\begin{equation}
\dot{t}=\frac{E_{\gamma}}{f(r)},
\label{eqn:first_int_t}
\end{equation}
\begin{equation}
\dot{\phi}=\frac{L_{\gamma}}{r^2},
\label{eq2}
\end{equation}
here, $E_{\gamma}$ and $L_{\gamma}$ are the energy and angular momentum of an incoming test particle.
The constraint on null geodesics for the line element given by Eqn.(\ref{BH1}) therefore reads as,
\begin{equation}
- f(r) \hspace{0.1cm} \dot{t}^2 + \frac{1}{f(r)} \hspace{0.1cm}\dot{r}^2 + r^2 \hspace{0.1cm} \dot{\phi}^2 = 0 .
\label{eq3}
\end{equation}
Further, using Eqn.(\ref{eqn:first_int_t}) and Eqn.(\ref{eq2}), the radial velocity in the equatorial plane is given as,
\begin{equation}
\dot{r}^2= E^{2}_{\gamma} - \frac{f(r)}{r^2}  \hspace{0.1cm}L^{2}_{\gamma}.
\label{eq4}
\end{equation}
Comparing Eqn.(\ref{eq4}) with $\dot{r}^2 + V(r) = 0$, the effective potential for null geodesics
can be defined as,
\begin{equation}
V(r) = \frac{f(r)}{r^2} \hspace{0.1cm} L^{2}_{\gamma} - E^{2}_{\gamma}.
\label{eq5}
\end{equation}
\subsection{Radial geodesics}
\begin{figure}
\includegraphics[width=8cm,height=6cm]{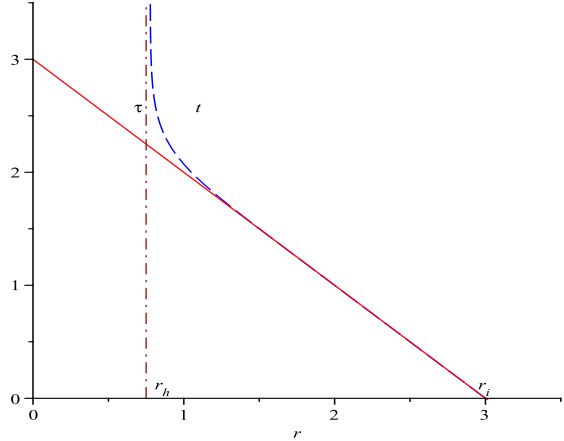}
\caption{Radial motion of an incoming massless test particle for proper time $\tau$ (solid line) and the coordinate time $t$ (dashed line) with $E_{\gamma} = 1$, $\epsilon = 0.1$ and $m = 1$. Here $r_h$ and $r_i$ on horizontal axis represent the position of event horizon and the initial position of the test particle respectively.}
\label{radial}
\end{figure}

\noindent The radial geodesics correspond to the trajectories for incoming photons with zero angular momentum (i.e. $L_{\gamma}$ = 0).
For radial null geodesics, the radial velocity given in Eqn.(\ref{eq4}) reduces to,
\begin{eqnarray}
\dot{r} = \pm \hspace{0.2cm} E_{\gamma},
\label{eq7}
\end{eqnarray}
on integrating Eqn.(\ref{eq7}), we obtain,
\begin{equation}
\tau = \frac{r}{E_{\gamma}} + constant.
\label{radial_tau}
\end{equation}
Eqn.(\ref{radial_tau}) shows that the radial coordinate depends only on energy $E_{\gamma}$ and is graphically presented in Fig.(\ref{radial}). One can notice from Fig.(\ref{radial}) that the behaviour of the proper time experienced by photons in this case is similiar to the Schwarzschild case.
In order to obtain the coordinate time $t$ as a function of radial distance $r$, dividing Eqn.(\ref{eq7}) by Eqn.(\ref{eqn:first_int_t}) leads to,
\begin{equation}
\frac{dt}{dr} = \pm \hspace{0.2cm} \frac{1}{f(r)} ,
\label{eq8}
\end{equation}
which can be further integrated as,
\begin{equation}
t(r) = \int_{r_i}^{r} \frac{dr}{\left[1 - \frac{4  \hspace{0.1cm}m}{r  \hspace{0.1cm}\sqrt{\pi}}  \hspace{0.1cm}\gamma\left(\frac{3}{2}, \frac{r^2}{4  \hspace{0.1cm}\epsilon}\right)\right]}.
\label{tr_relation}
\end{equation}
For the t-r relationship given by Eqn.(\ref{tr_relation}) as depicted in Fig.(\ref{radial}), one can easily notice that a massless test particle crosses the event horizon in a finite proper time ($\tau$) while it approaches asymptotically the event horizon in terms of the coordinate time (t).
\subsection{Non-radial geodesics}
\begin{figure}
\centerline{
\includegraphics[scale=0.8]{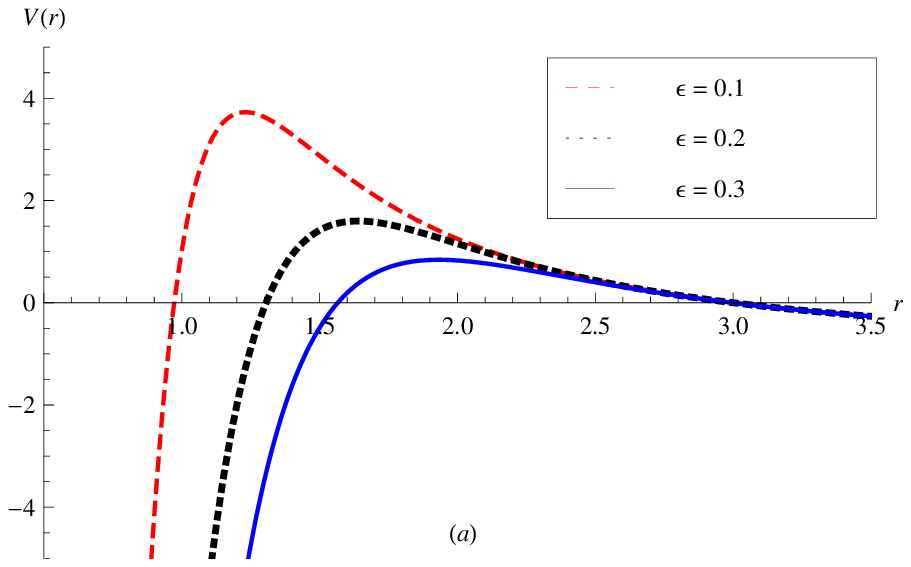}
\includegraphics[scale=0.8]{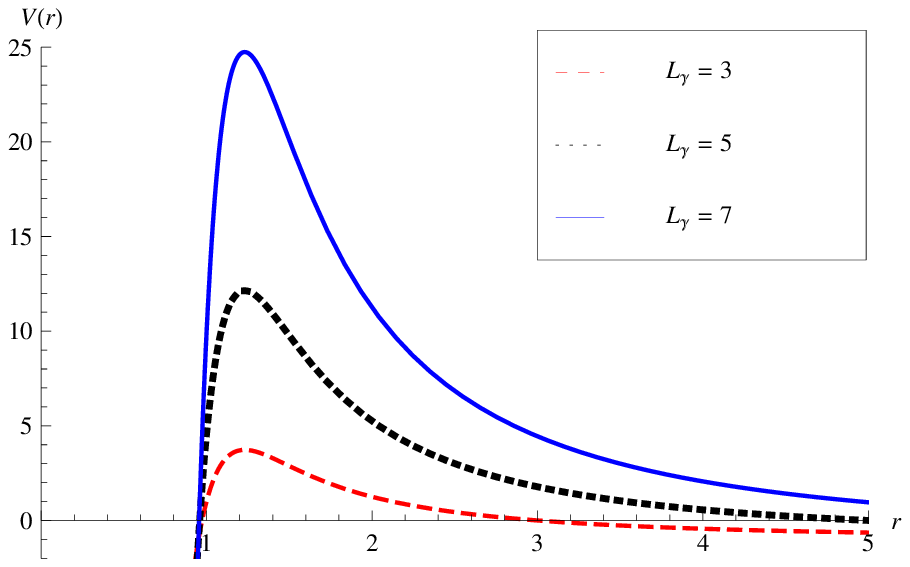}}
\caption{Variation of effective potential with (a) non-commutative parameter  $\epsilon$ (m = 2, $L_{\gamma} = 3$ and $E_{\gamma} = 1$), (b) angular momentum $L_{\gamma}$ ($\epsilon = 0.1$, m = 2 and $E_{\gamma} = 1$).}
\label{f3}
\end{figure}

\noindent Now, let us discuss the null geodesics with angular momentum (i.e. $L_{\gamma} \neq$ 0) i.e. non-radial geodesics.
The effective potential for non-radial null geodesics is given by,
\begin{equation}
V(r) = \frac{f(r)}{r^2} \hspace{0.1cm} L_{\gamma}^{2} - E_{\gamma}^{2}.
\label{eq9}
\end{equation}
In Fig.\ref{f3}(a), the effective potential has been presented for different values of $\epsilon$. It can be seen that the height of potential decreases with the increasing values of parameter $\epsilon$. The effective potential shows a maxima which corresponds to an unstable circular orbit. It is also observed that with the increase in the value of non-commutative parameter, the peak is shifting towards the right, which signifies the shifting of circular orbit away from the central object. The variation of effective potential with angular momentum is shown in Fig.\ref{f3}(b) and one can note that the peaks of potential increases with increasing values of angular momentum.
In case of null geodesics, $E_{\gamma}$ and $L_{\gamma}$ do not have direct physical significance as such. However, their ratio $L_{\gamma}/E_{\gamma}$ i.e. impact parameter (D) plays a similar role as that of the energy and angular momentum of the incoming test particle in case of timelike geodesics \cite{har}.
\subsection{Analysis of photon orbit}
\begin{figure}[h]
\centerline{
\includegraphics[width=6cm,height=5cm]{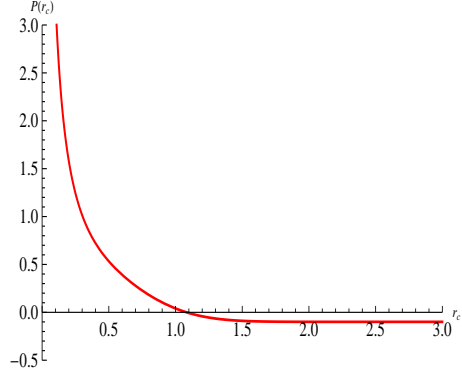}}
\caption{Variation of $P(r)$ with $r$ for $m = 1$ and $\epsilon = 0.1$. Here, only one real root exists.}
\label{function1}
\end{figure}

\begin{figure}[h]
\centerline{
\includegraphics[width=6cm,height=5cm]{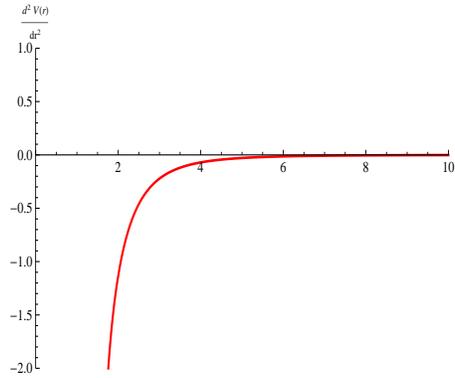}}
\caption{Unstability condition of massless particles for $m = 1$, $\epsilon = 0.1$ and $L = 3$.}
\label{unstability}
\end{figure}

\begin{figure}[h]
\centerline{
\includegraphics[width=6cm,height=5cm]{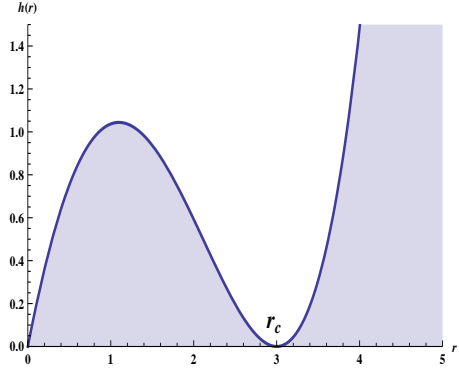}}
\caption{Variation of $h(r)$ with $r$ for $\epsilon = 0.1$, m = 1 and $D = D_c = 5.19$. Here, shaded region is the allowed region for the motion of photons.}
\label{f1}
\end{figure}

\begin{figure}[h]
\centerline{
\includegraphics[width=6cm,height=6cm]{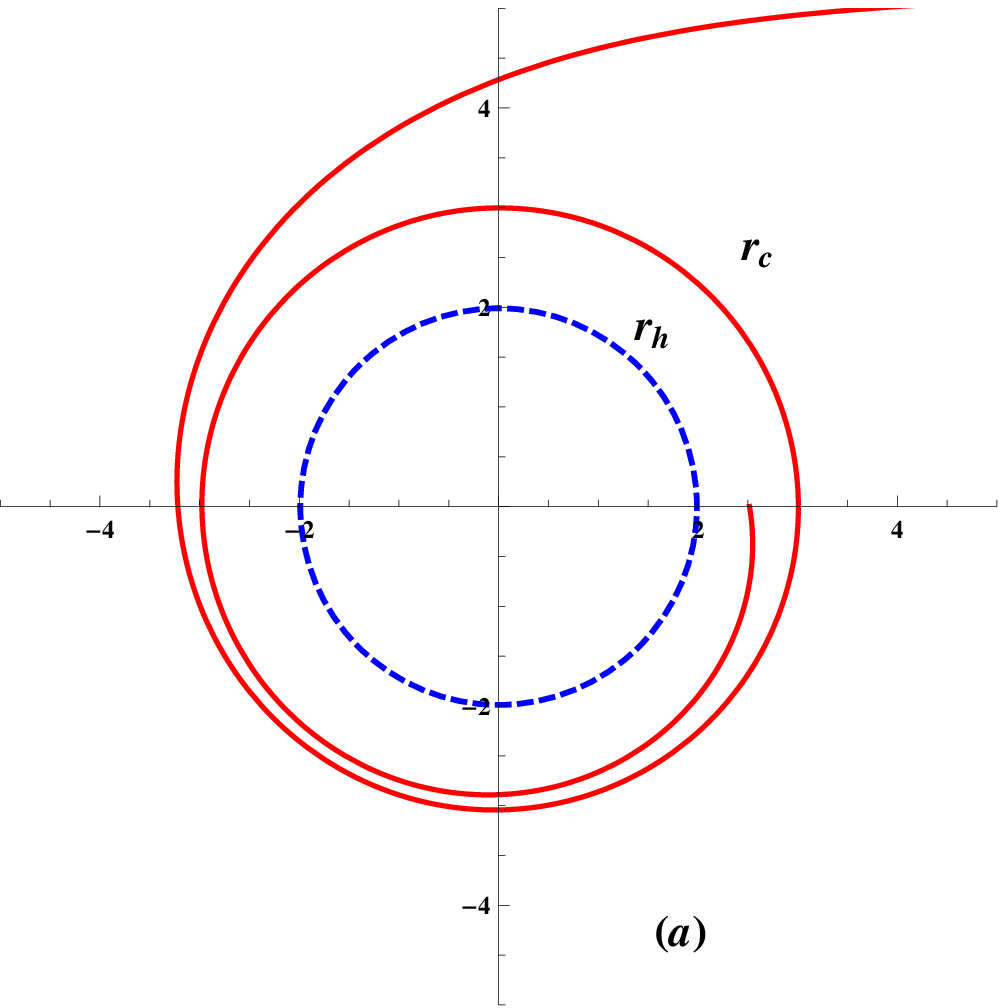}
\includegraphics[width=6cm,height=6cm]{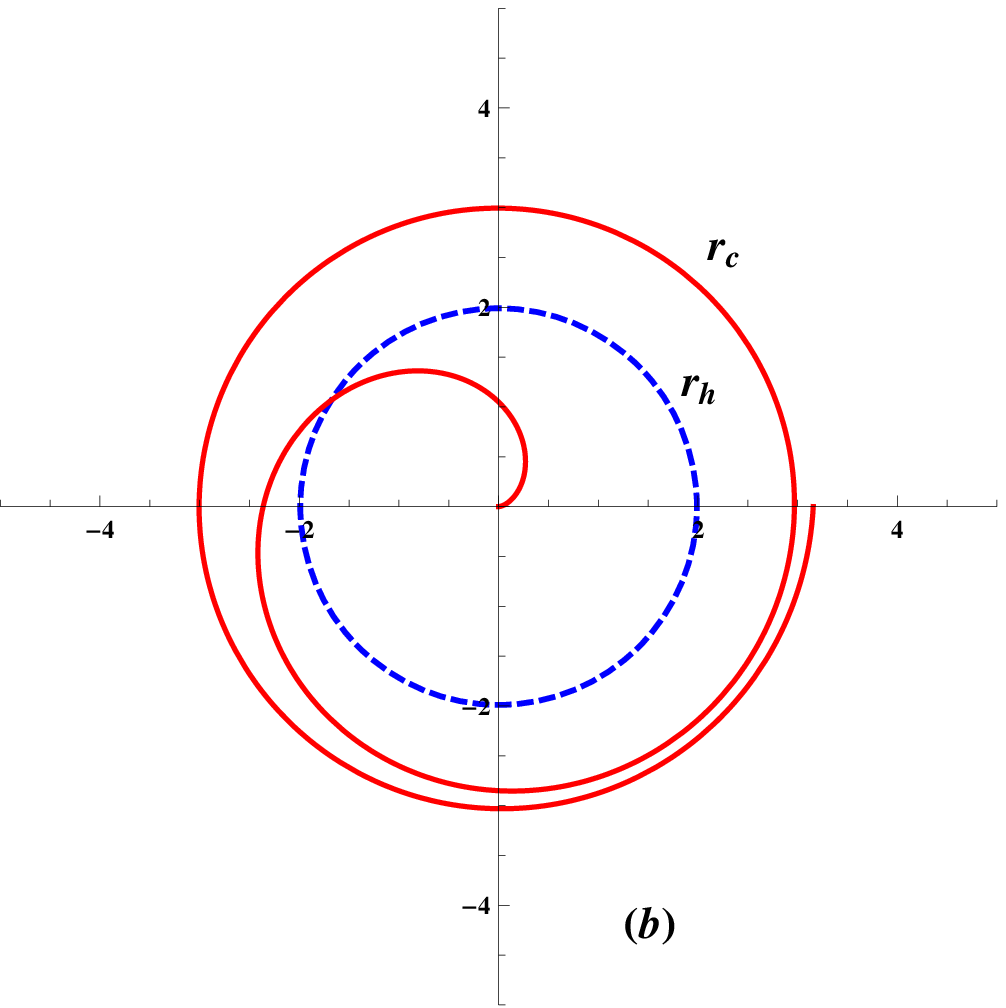}}
\caption{The circular orbit (a) for a photon starting from infinity, (b) a bound orbit spiralling around the circular radius $r_c = 2.99$ starting form the singularity. Here, $\epsilon = 0.1$, m = 2, $D = D_c =5.19$ and $r_h$ is the radius of event horizon.}
\label{co}
\end{figure}

\begin{figure}[h]
\centerline{
\includegraphics[width=6cm,height=5cm]{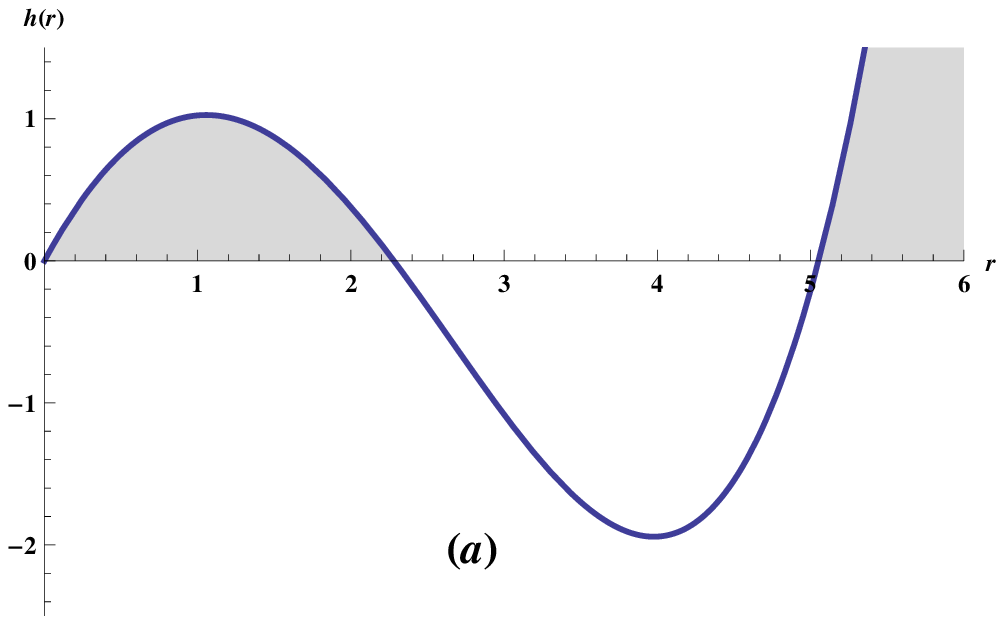}
\includegraphics[width=5cm,height=5cm]{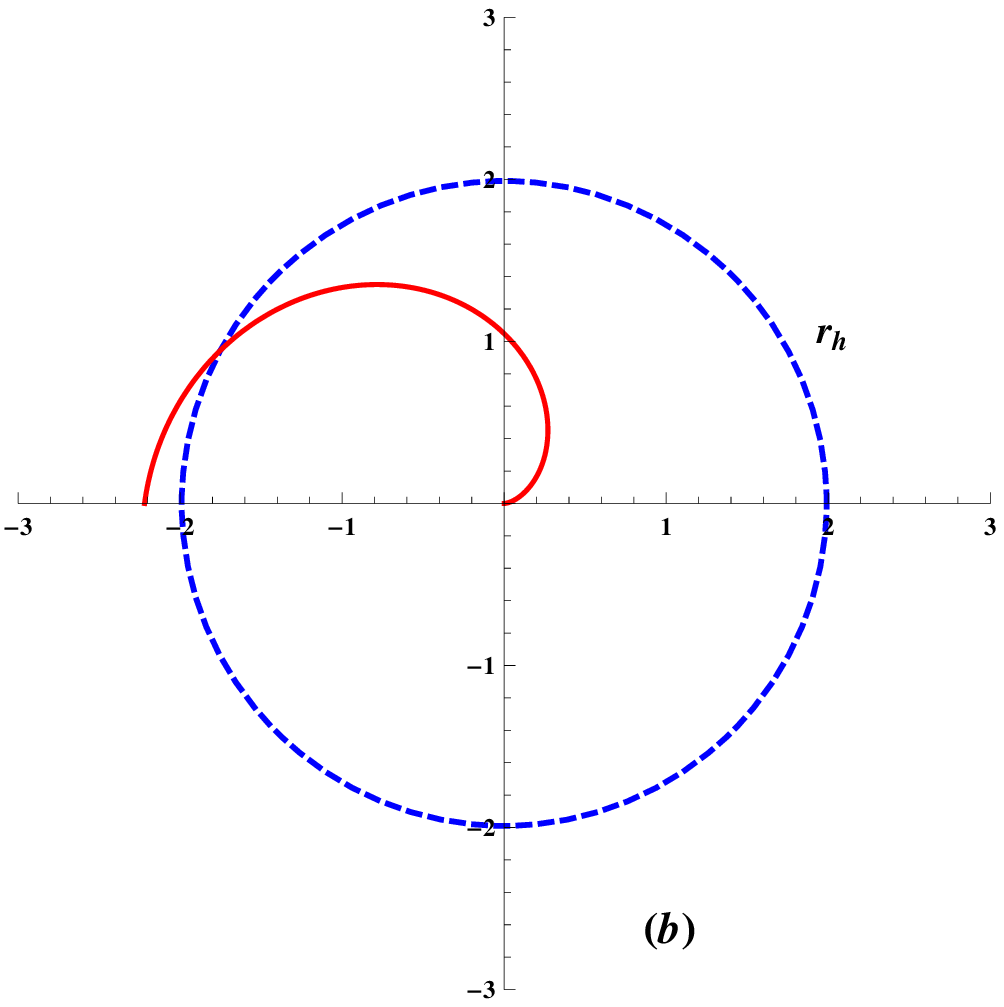}
\includegraphics[width=4.7cm,height=5cm]{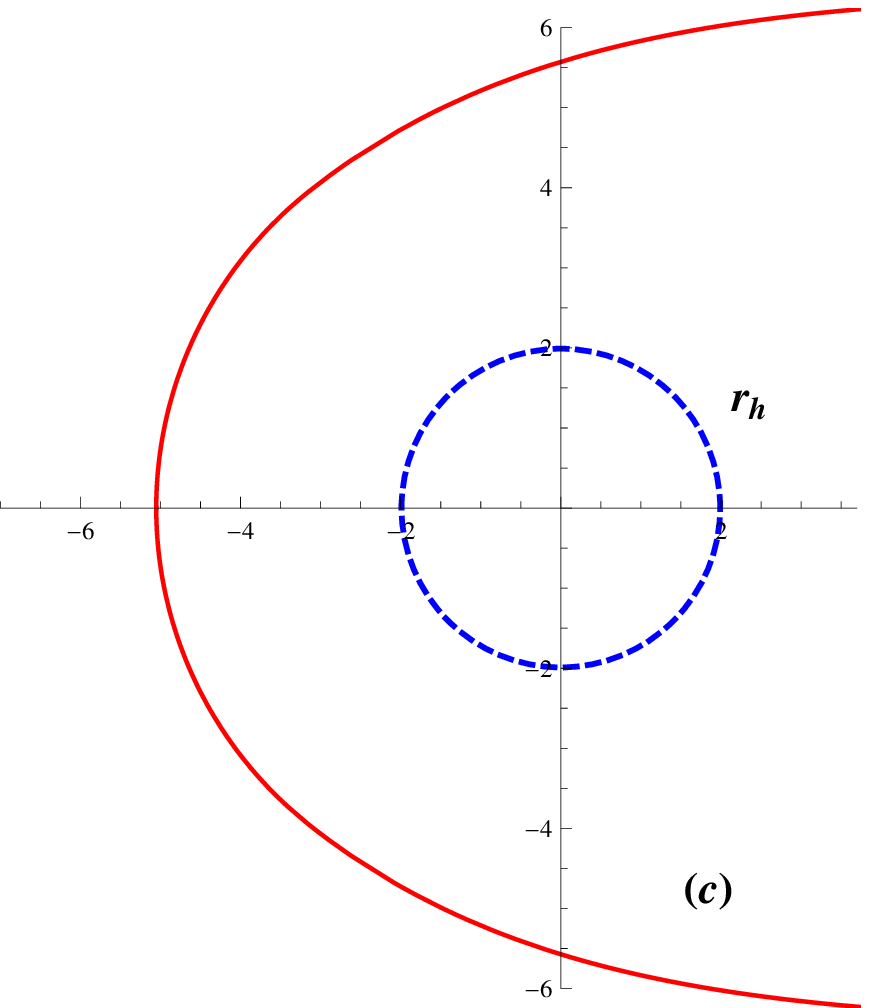}}
\caption{(a) Variation of $h(r)$ with $r$ for $\epsilon = 0.1$, m = 2 and $D = 6.5 > D_c$. Here, the shaded region is the allowed region for the motion of photons, (b) solid line represents a bound orbit, (c) solid line represents the fly-by orbit for the null geodesics approaching the BH from infinity and again flying back to infinity. Here, $r_h$ is the radius of event horizon.}
\label{o2}
\end{figure}

\begin{figure}[h]
\centerline{
\includegraphics[width=6cm,height=5cm]{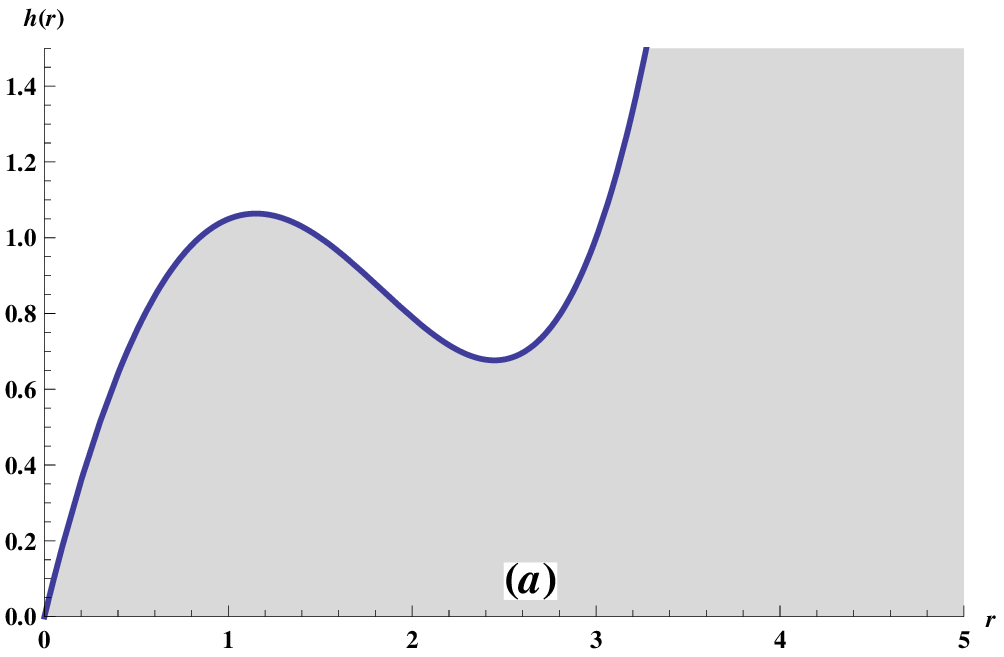}
\includegraphics[width=5cm,height=5cm]{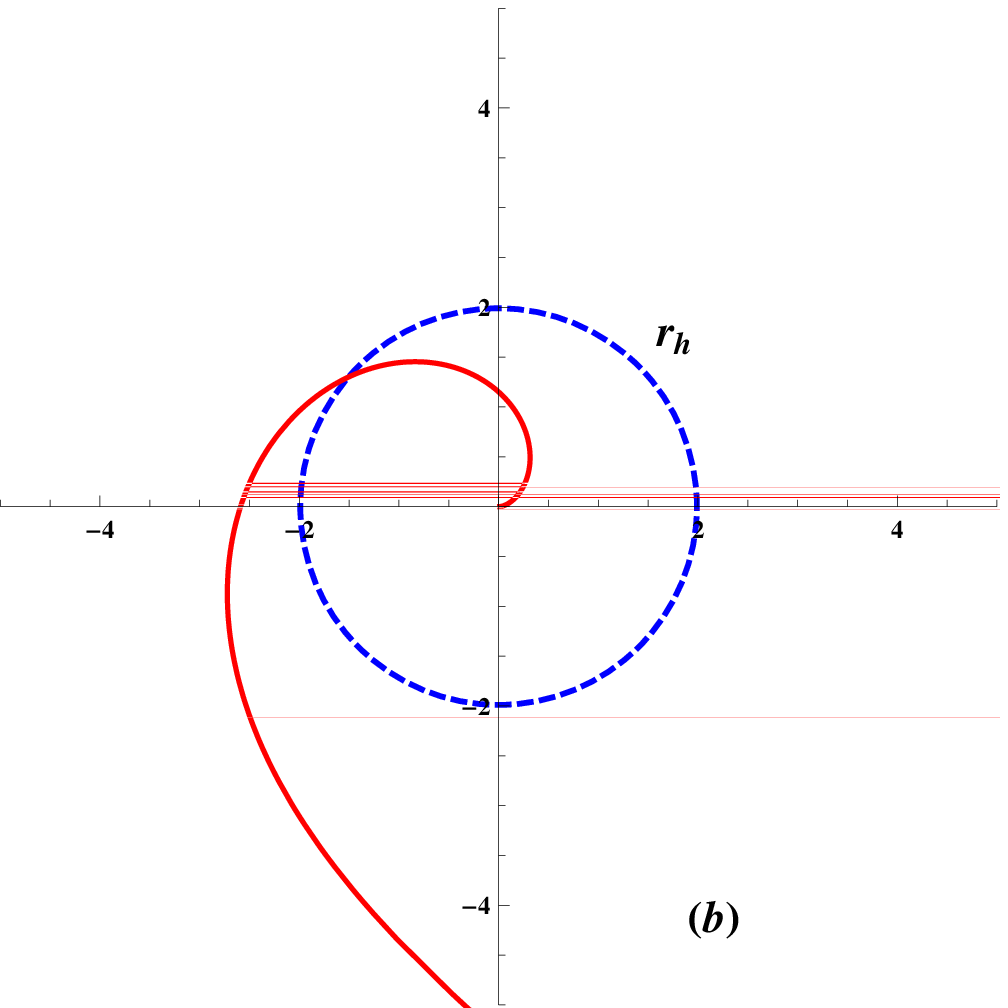}}
\caption{(a) Variation of $h(r)$ with $r$ for $\epsilon = 0.1$, m = 2 and $D = 4.5 < D_c$ . The shaded region is the allowed region for the motion of photons, (b) solid line represents the terminating escape orbit for the null geodesics approaching the BH from infinity. Here, $r_h$ is the radius of event horizon.}
\label{o3}
\end{figure}

\noindent To analyse the motion of photons, the corresponding orbit equation can be obtained by using Eqn.(\ref{eq2}) and Eqn.(\ref{eq4}) as,
\begin{equation}
\left(\frac{dr}{d\phi}\right)^2 = \frac{r^4}{D^2} - r^2 f(r).
\label{eq10}
\end{equation}
This expression can be re-expressed as,
\begin{equation}
\left(\frac{dr}{d\phi}\right)^2 = h(r),
\label{eq11}
\end{equation}
with,
\begin{equation}
h(r) = \frac{r^4}{D^2} - r^2 f(r).
\end{equation}
In view of orbit Eqn.(\ref{eq10}), it is clear that the geometry of the null geodesics depends on the nature of the roots of the function $h(r) = 0$. All the possible orbits for null geodesics in view of the different values of impact parameter $D$ may now be discussed accordingly.\\
\textbf{Case A : For D = Dc (circular photon orbit)}\\
The condition of circular orbit i.e.,
\begin{equation}
\dot{r} = 0 \Rightarrow V(r) = 0,
\label{eq12}
\end{equation}
along with $V^{'}(r) = 0$\footnote{where $'$ denotes the differentiation w.r.t. r .} leads to the impact parameter for unstable circular orbit (i.e. $D_c$),
\begin{equation}
D^2_{c} = \frac{r^2_{c}}{f(r_c)} = \frac{r^2_{c}}{ 1 - \frac{4 \hspace{0.1cm} m}{r_c \hspace{0.1cm} \sqrt{\pi}} \hspace{0.1cm} \gamma\left(\frac{3}{2}, \frac{r_{c}^2}{4 \hspace{0.1cm} \epsilon}\right)},
\label{impact}
\end{equation}
and the radius of unstable circular orbit (i.e. $r_c$),
\begin{eqnarray}
P(r_c) \equiv r_{c}^2 \hspace{0.1cm} e^{\frac{-r_{c}^2}{4\hspace{0.1cm}\epsilon}} + \frac{6 \hspace{0.1cm}\sqrt{\pi} \hspace{0.1cm} \epsilon^{3/2}}{r_{c}}\hspace{0.1cm}\gamma\left(\frac{3}{2}, \frac{r_{c}^2}{4\hspace{0.1cm} \epsilon}\right) - \frac{\pi\hspace{0.1cm} \epsilon^{3/2}}{m} = 0.
\label{poly1}
\end{eqnarray}
One can observe from Eqns.(\ref{impact}) and (\ref{poly1}) that the impact parameter and the radius of unstable circular orbit are independent of the energy ($E_{\gamma}$) and angular momentum ($L_{\gamma}$) of the masless particles and only depends on the mass and non-commutative parameter of a NCBH spacetime. One can obtain the radius of unstable circular orbit numerically by analysing the behavior of $P(r)$ with $r$ as depicted in Fig.(\ref{function1}). The Eqn.(\ref{poly1}) has at least one real root where it cuts the $r$ axis (see Fig.(\ref{function1})). Hence, it is possible to have a bound orbit for the test particles. In other words, the massless test particles can be trapped by the gravitational field of NCBH.
The unstability of the circular orbit is given by,
\begin{equation}
\frac{d^2 V(r)}{dr^2}_{r= r_c} < 0.
\label{eq14}
\end{equation}
The unstability condition of circular orbit of massless particles is presented visually in Fig.(\ref{unstability}), which shows that the circular orbits will always be unstable.
Further, it is not possible to solve the orbit equation given by Eqn.(\ref{eq11}) analytically. So, we reduces the lower incomplete gamma function into upper incomplete gamma function and then the possible orbits are presented numerically for different values of impact parameter. By using the upper incomplete gamma function, the expression given by Eqn.(\ref{poly1}) for radius of unstable circular orbit looks like,
\begin{eqnarray}
r_c = 3 \hspace{0.1cm} m \left(1 - \frac{m}{\sqrt{\pi \epsilon}} \hspace{0.1cm} e^{\frac{-m^2}{\epsilon}}\right).
\label{approx radius}
\end{eqnarray}
One can see from Eqn.(\ref{approx radius}), the radius of unstable circular orbit in case of NCBH is smaller than the SBH ($r_c = 3m$). This implies that due to the presence of noncommutativity, the gravitational field of NCBH becomes weaker than SBH spacetime in GR.
The variation of function $h(r)$ with $r$ is presented in Fig.(\ref{f1}).
The possible orbit of photon for the allowed regions [see Fig.(\ref{f1})] is numerically presented in Fig.(\ref{co}).\\
{\bf The Time Period}:\\
The time period for circular orbits can be calculated for proper time as well as coordinate time with $\phi = 2 \pi$ \cite{RefF1}. For NCBH, using Eqns.(\ref{eqn:first_int_t}) and (\ref{eq2}) and integrating for one complete time period, the time period in terms of proper time and coordinate time comes out respectively as below,
\begin{equation}
T_{\tau} = \frac{2 \hspace{0.1cm} \pi \hspace{0.1cm} r^{2}_{c}}{L_{\gamma}},
\end{equation}
and
\begin{equation}
T_{t} = \frac{2 \hspace{0.1cm} \pi \hspace{0.1cm} r_c}{\sqrt{f(r_c)}}.
\end{equation}
One can compute $T_{\tau}$ and $T_t$ for the Schwarzschild BH by considering the limit $\epsilon \rightarrow 0$. By observing the representative plots of the time periods, it is clear that the time periods for the NCBH are lower than the corresponding Schwarzschild case, which is consistent with the fact that due to the non-commutative parameter the gravitational attraction in NCBH case decreases.\\
\begin{figure}[h]
\centerline{
\includegraphics[width=6cm,height=5cm]{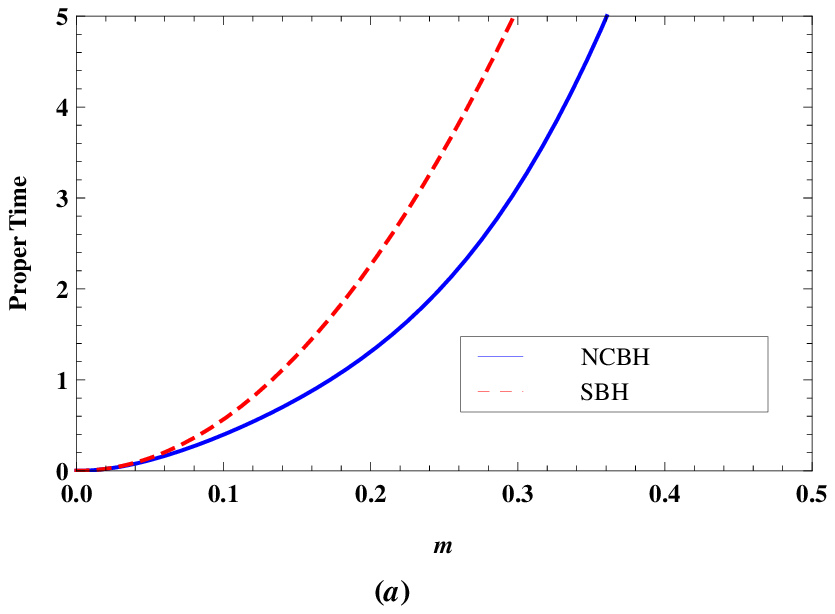}
\includegraphics[width=6cm,height=5cm]{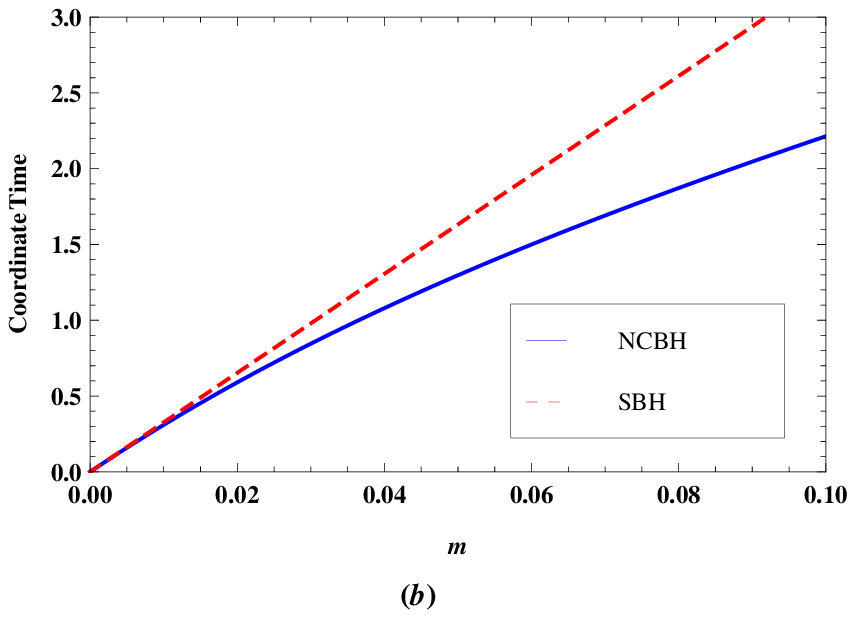}}
\caption{Variation of (a) proper time with $m$ and (b) coordinate time with $m$. Here, $\epsilon =0.1$ and $L_{\gamma} = 1$.}
\label{time period}
\end{figure}
\newpage
{\bf Case B: For $D > D_c$}\\
Here, the value of impact parameter $D = 6.5$, which is larger than the impact parameter for the case of unstable circular orbit. In Fig.(\ref{o2}), the allowed region for the motion of photon and the corresponding types of orbit are presented. There are two possible orbits, one is terminating bound and the another one is a fly-by orbit. One can see from the Fig.\ref{o2}(c) that the photon comes from infinity and after approaching the turning point it again flies back towards infinity.
\\
{\bf Case C: For $D < D_c$}\\
Now, with impact parameter $D = 4.5$ which is smaller than the impact parameter for the case of unstable circular orbit, the allowed region for the motion of photon and the corresponding terminating escape orbit is shown in Fig.(\ref{o3}).
\section{Frequency shift of photons emitted by massive particles}
\begin{figure}[h]
\centerline{
\includegraphics[width=6cm,height=5cm]{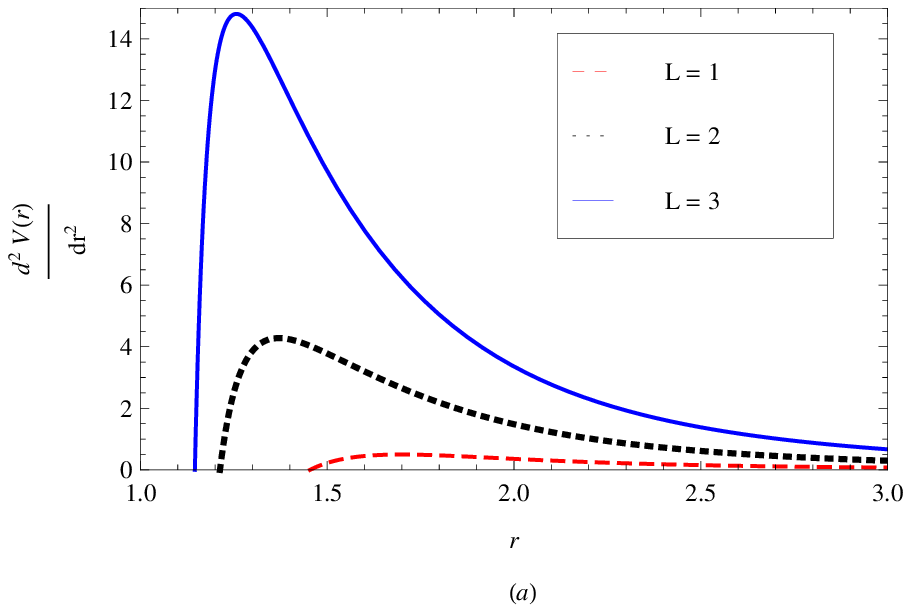}
\includegraphics[width=6cm,height=5cm]{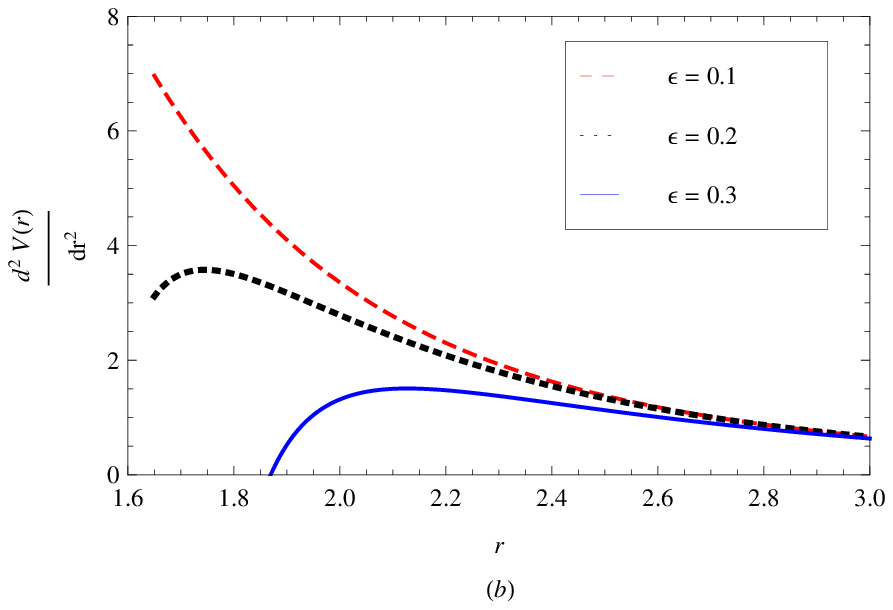}}
\caption{The variation of stability condition of massive particles with (a) angular momentum for $m = 1$ and $\epsilon = 0.1$, (b) non-commutative parameter for $m = 1$ and $L = 3$.}
\label{stability}
\end{figure}

\noindent Let us consider the photons with four-momentum, parametrised by $k^{\mu} = (k^{t}, k^{r}, k^{\theta}, k^{\phi})$, which move along the null geodesics ($k^{\mu}k_{\mu} = 0$). With this, the null geodesic constraint equation reads as,
\begin{equation}
- g_{tt} (k^{t})^2 + g_{rr} (k^{r})^2 + g_{\phi \phi} (k^{\phi})^2 = 0,
\end{equation}
where, $g_{tt} = f(r)$, $g_{rr} = f(r)^{-1}$ and $g_{\phi \phi} = r^2$. The conserved quantities for these photons are therefore,
\begin{eqnarray}
E_{\gamma} = f(r)\hspace{0.2cm} k^{t},\\
L_{\gamma} = r^2 \hspace{0.2cm}k^{r}.
\label{E and L photon}
\end{eqnarray}
In order to compute red/blue shifts that emitted photons by massive particles experience while traveling along null geodesics towards an observer located far away from their source, we follow the approach as mentioned in \cite{alf15, bec}.\\
The general expression for the frequency of a photon measured by an observer with proper 4-velocity $u^{\mu}_{c}$ at point $P_c$ reads,
\begin{equation}
w_c = - k_{\mu} u^{\mu}_{c}\vert_{P_c},
\end{equation}
where, the index c refers to the emission (e) and/or detection (d) at the corresponding spacetime point $P_c$.
Thus, the frequency of the light signals measured by the comoving observer at the emission point (e) is,
\begin{equation}
w_e = -(k_{\mu} u^{\mu})\vert_e,
\end{equation}
whereas the frequency detected far away from the source is,
\begin{equation}
w_d = -(k_{\mu} u^{\mu})\vert_d,
\end{equation}
where, the four velocities of the emitter and the detector are given respectively as below,
\begin{eqnarray}
u^{\mu}_e = (u^t, u^r, u^{\theta}, u^{\phi}),\vert_e\\,
u^{\mu}_d = (u^t, u^r, u^{\theta}, u^{\phi}). \vert_d .
\end{eqnarray}
Now, the four velocities reads in the case when the detector is situated at a large distance from the source ($r \rightarrow \infty$) as follows,
\begin{equation}
u^{\mu}_d = (1, 0, 0, 0).
\end{equation}
On the other hand, a photon which is emitted or detected at point $P_c$ possesses a four momentum $k^{\mu}_c = (k^t, k^r, k^{\theta}, k^{\phi})_c$.\\
Thus, the frequency shift associated with the emission and detection of photons is given by,
\begin{eqnarray}
1 + z = \frac{w_e}{w_d} , \nonumber \\
= \frac{(E_{\gamma} u^t - L_\gamma u^{\phi} - g_{rr} u^r k^r - g_{\theta \theta} u^{\theta} k^{\theta})\vert_e}{(E_{\gamma} u^t - L_\gamma u^{\phi} - g_{rr} u^r k^r - g_{\theta \theta} u^{\theta} k^{\theta})\vert_d}.
\end{eqnarray}
In the case of circular and equatorial plane ($u^r = u^{\theta} =0$), the expression of the red/blue shift becomes,
\begin{eqnarray}
1 + z = \frac{(E_{\gamma} u^t - L_\gamma u^{\phi})\vert_e}{(E_{\gamma} u^t - L_\gamma u^{\phi})\vert_d} ,\nonumber \\
= \frac{u^{t}_e - D_e u^{\phi}_e}{u^{t}_d - D_d u^{\phi}_d},
\label{shift1}
\end{eqnarray}
here, $D (\equiv L_{\gamma}/E_{\gamma})$ is the impact parameter.\\
Further, we consider the kinematic red/blue shift of photons either side of the central value of impact parameter i.e. $D = 0$. The expression for red shift corresponding to a photon emitted by a static particle located at $D = 0$ reads as,
\begin{equation}
1 + z_c = \frac{u^{t}_e}{u^{t}_d}.
\label{shift2}
\end{equation}
However, the expression of kinematical red shift ($z_{kin}$) can be obtained by subtracting Eqn.(\ref{shift2}) from Eqn.(\ref{shift1}),
\begin{equation}
z_{kin} \equiv z -z_c = \frac{u^{t}_e u^{\phi}_d D_d - u^{t}_d u^{\phi}_e D_e}{u^{t}_e (u^{t}_d - D_d u^{\phi}_d)}.
\label{kinetic shift}
\end{equation}
From the expression (\ref{kinetic shift}), it follows that the apparent impact parameter $D$ must also be maximized; this quantity can be calculated from the geodesic equation of the photons (or, equivalently, from the relation $k^{\mu} k_{\mu} = 0$ (with $k^{r} = 0$ and $k^{\theta} = 0$) as given below,
\begin{equation}
D_{\pm} = \pm \hspace{0.2cm} \sqrt{\frac{g_{\phi \phi}}{g_{tt}}},
\end{equation}
or,
\begin{equation}
D_{\pm} = \pm \hspace{0.2cm} \frac{r}{\sqrt{f(r)}}.
\label{impact_b}
\end{equation} 
Here these two values ($D_{+}$ and $D_{-}$) correspond to the emitter or detector position, since this quantity is preserved along the null geodesic trajectories of the photons, (i.e. $D_e = D_d$) that respectively give rise to two different shifts, $z_1$ and $z_2$, of the emitted photons as expressed below,
\begin{eqnarray}
z_1 = \frac{u^{t}_e u^{\phi}_d D_{d_{-}} - u^{t}_d u^{\phi}_e D_{e_{-}}}{u^{t}_d (u^{t}_d - u^{\phi}_d D_{d_{-}})},\\
z_2 = \frac{u^{t}_e u^{\phi}_d D_{d_{+}} - u^{t}_d u^{\phi}_e D_{e_{+}}}{u^{t}_d (u^{t}_d - u^{\phi}_d) D_{d_{+}}}.
\label{red_blue shift1}
\end{eqnarray}
In order to have a close expression for the gravitational
red/blue shifts experienced by the emitted photons, one would express the required quantities in terms of the NCBH metric. \\
Now, by using the geodesic equations and constraint equation (i.e $\ddot{x^a}+\Gamma^{a}_{bc}\dot{x}^{b}\dot{x}^{c}=0 $ and  $g_{{a}{b}}\dot{x}^a \dot{x}^b=-1$), the four velocity components of massive test particle in equatorial plane can be calculated as,
\begin{eqnarray}
u^{t} = \frac{E}{f(r)},\nonumber \\
u^{\phi} = \frac{L}{r^2},
\label{first intgrl}
\end{eqnarray}
where $E$ and $L$ correspond to the conserved total energy and the conserved angular momentum of a massive test particle respectively.
The effective potential thus acquires the following form,
\begin{equation}
V_{eff} = 1 -\frac{E^2}{f(r)} + \frac{L^2}{r^2}.
\label{potential_massive}
\end{equation}
Again from the condition of circular orbit, one can find the general expressions for the constants of motion as,
\begin{eqnarray}
E = f(r_s) \hspace{0.2cm} \sqrt{\frac{- 2}{r_s \hspace{0.2cm}A(r_s) - 2 \hspace{0.2cm}f(r_s)}}, \label{energy1} \\
L = \sqrt{\frac{ - r_s^{3}\hspace{0.2cm} A(r_s)}{r_s \hspace{0.2cm} A(r_s) - 2 \hspace{0.2cm}f(r_s)}},
\label{angular1}
\end{eqnarray}
where, $A(r) = \frac{2 m r_s\hspace{1mm} e^{-\frac{r_{s}^2}{4 \epsilon}}}{\pi \epsilon^{3/2}} + \frac{4 m \gamma\left(\frac{3}{2}, \frac{r_{s}^2}{4 \epsilon}\right)}{\sqrt{\pi} r^2}$ and $ f(r_s) =1 - \frac{4 m}{r_s \sqrt{\pi}} \gamma\left(\frac{3}{2}, \frac{r_{s}^2}{4 \epsilon}\right)$. 
Here $E$ and $L$ should be real so, $r_s A(r_s) < 2 f(r_s)$ i.e.,
\begin{equation}
2m r^{2}_{s} e^{-r^{2}_{s}/4 \epsilon} + \frac{4 m}{\sqrt{\pi} r_s} \gamma\left(\frac{3}{2}, \frac{r_{s}^2}{4 \epsilon}\right) -2 - \frac{8 m }{\sqrt{\pi} r_s} \gamma\left(\frac{3}{2}, \frac{r_{s}^2}{4 \epsilon}\right) < 0,
\end{equation}
or,
\begin{equation}
1 + \frac{2}{\sqrt{\pi} \tilde{r_s}} \gamma\left(\frac{3}{2}, \frac{\tilde{r_s}^2}{4 \tilde{\epsilon}}\right) - \frac{\tilde{r_s}^2}{\pi\tilde{\epsilon}^{3/2} e^{-\frac{\tilde{r_s}^2}{4 \tilde{\epsilon}}}} < 0,
\end{equation}
where, $\tilde{r_s} = \frac{r_s}{m}$ and $\tilde{\epsilon} = \frac{\epsilon}{m^2}$.
The radius of stable circular orbit $r_s$ is given by the following polynomial,
\begin{eqnarray}
Q(r_s)\equiv r_{s}^4 e^{-r_{s}^2/4 \epsilon} + r_{s}^2 e^{-r_{s}^2/4 \epsilon} L^2 + 2 r_{s} \sqrt{\pi} \epsilon^{3/2} \gamma\left(\frac{3}{2}, \frac{r_{s}^2}{4 \epsilon}\right) \nonumber \\
 + \frac{6 L^2}{ r_{s}} \sqrt{\pi} \epsilon^{3/2} \gamma\left(\frac{3}{2}, \frac{r_{s}^2}{4 \epsilon}\right) - \frac{L^2 \pi \epsilon^{3/2}}{m} = 0,
\label{poly2}
\end{eqnarray}
which after a parametrization in terms of $\tilde{L} = L/m$ can be re-expressed as below,
\begin{eqnarray}
Q(r_s)\equiv \tilde{r_s}^4 e^{-\tilde{r_s}^2/4 \tilde{\epsilon}} + \tilde{r_s}^2 e^{-\tilde{r_s}^2/4 \tilde{\epsilon}} L^2 + 2 \tilde{r_s} \sqrt{\pi} \tilde{\epsilon}^{3/2} \gamma\left(\frac{3}{2}, \frac{\tilde{r_s}^2}{4 \tilde{\epsilon}}\right) \nonumber \\
 + \frac{6 \tilde{L}^2}{\tilde{r_s}} \sqrt{\pi} \tilde{\epsilon}^{3/2} \gamma\left(\frac{3}{2}, \frac{\tilde{r_s}^2}{4 \tilde{\epsilon}}\right) - \frac{\tilde{L}^2 \pi \tilde{\epsilon}^{3/2}}{m} = 0.
\label{poly3}
\end{eqnarray}
The general expression from the stability condition of circular orbit i.e ($V^{''}_{eff} > 0$) comes out as,
\begin{equation}
V^{''}_{eff} =\frac{2 \hspace{0.1cm}B(r_s)\hspace{0.1cm} f(r_s) - 4 \hspace{0.1cm}A^{2}(r_s)}{f(r_s)\hspace{0.1cm} [2 \hspace{0.1cm}f(r_s) - r\hspace{0.1cm} A(r_s) ]} +  \frac{6\hspace{0.1cm} L^2}{r_{s}^4},
\label{stability condition}
\end{equation}
or using Eqn.(\ref{angular1}), the above equation can be written as,
\begin{equation}
V^{''}_{eff} = \frac{- 2 \hspace{0.1cm}r_{s}\hspace{0.1cm} B(r_s)\hspace{0.1cm} f(r_s)+ 4\hspace{0.1cm} r_{s}\hspace{0.1cm} A^{2}(r_s) - 6\hspace{0.1cm} f(r_s) \hspace{0.1cm}A(r_s)}{r \hspace{0.1cm}f(r_s) \left[r_s\hspace{0.1cm} A(r_s) - 2\hspace{0.1cm} f(r_s)\right]},
\label{stability condition}
\end{equation}
where, $B(r_s) =-\frac{ m\hspace{0.1cm} r_{s}^2\hspace{1mm} e^{-\frac{r^2}{4 \hspace{0.1cm}\epsilon}}}{\pi \hspace{0.1cm}\epsilon^{5/2}} - \frac{8\hspace{0.1cm} m\hspace{0.1cm} \gamma\left(\frac{3}{2}, \frac{r_{s}^2}{4 \hspace{0.1cm}\epsilon}\right)}{\sqrt{\pi} \hspace{0.1cm}r_{s}^3}$.
The stability condition is graphically presented in Fig.(\ref{stability}). One can observe that on increasing the value of angular momentum, the peaks of stability shift in upward direction. Whereas, the peaks in the plot shifts in downward direction on increasing the value of the non-commutative parameter.
From the stability condition, using Eqn.(\ref{stability condition}), one can write,
\begin{equation}
\frac{r_s \hspace{0.1cm} B(r_s) \hspace{0.1cm} f(r_s) - 2 \hspace{0.1cm} r_s \hspace{0.1cm} A^{2}(r_s) + 3 \hspace{0.1cm} f(r_s) \hspace{0.1cm} A(r_s)}{f(r_s)} > 0,
\end{equation}
which in turn leads the bound on radius as,
\begin{eqnarray}
\frac{m}{r^2_{s} \left( 2 \hspace{0.1cm} m  \hspace{0.1cm}e^{-\frac{r^2_{s}}{4  \hspace{0.1cm}\epsilon}} \hspace{0.1cm} \pi^{-1/2}  \hspace{0.1cm}\epsilon^{-1/2} \hspace{0.1cm} r_s - 2  \hspace{0.1cm}m + r_s\right)} \times \nonumber \\
\Big[
\Big(-\frac{ 2  \hspace{0.1cm}m  \hspace{0.1cm}e^{-\frac{r^2_{s}}{2 \hspace{0.1cm}\epsilon}}}{\pi^{3/2}  \hspace{0.1cm}\epsilon^3} - \frac{8 \hspace{0.1cm} m  \hspace{0.1cm}e^{-\frac{r^2_{s}}{2 \hspace{0.1cm} \epsilon}}}{\pi^{2}  \hspace{0.1cm}\epsilon^3} - \frac{e^{-\frac{r^2_{s}}{4  \hspace{0.1cm}\epsilon}}}{\pi \hspace{0.1cm} \epsilon^{5/2}}\Big) \hspace{0.1cm}r^6_{s} + \frac{2  \hspace{0.1cm}m  \hspace{0.1cm}e^{-\frac{r^2_{s}}{4 \hspace{0.1cm} \epsilon}}}{\pi \hspace{0.1cm}\epsilon^{5/2}} \hspace{0.1cm}  \hspace{0.1cm}r^5_{s} + 2  \hspace{0.1cm}\Big(\frac{3  \hspace{0.1cm}e^{-\frac{r^2_{s}}{4  \hspace{0.1cm}\epsilon}}}{\pi  \hspace{0.1cm}\epsilon^{3/2}} + \frac{14  \hspace{0.1cm}m  \hspace{0.1cm}e^{-\frac{r^2_{s}}{2 \hspace{0.1cm} \epsilon}}}{\pi^{3/2} \hspace{0.1cm} \epsilon^2} \Big)  \hspace{0.1cm}r^4_{s} \nonumber \\ - \frac{28  \hspace{0.1cm}m  \hspace{0.1cm}e^{-\frac{r^2_{s}}{4 \hspace{0.1cm} \epsilon}}}{\pi  \hspace{0.1cm}\epsilon^{3/2}}  \hspace{0.1cm}r^3_{s} 
- \Big(\frac{2  \hspace{0.1cm}e^{-\frac{r^2_{s}}{4  \hspace{0.1cm}\epsilon}}}{\sqrt{\pi  \hspace{0.1cm}\epsilon}} + \frac{2  \hspace{0.1cm}m \hspace{0.1cm} e^{-\frac{r^2_{s}}{2  \hspace{0.1cm}\epsilon}}}{\pi  \hspace{0.1cm}\epsilon} \Big) \hspace{0.1cm} r^2_{s}
+ \frac{24  \hspace{0.1cm}m  \hspace{0.1cm}e^{-\frac{r^2_{s}}{4 \hspace{0.1cm} \epsilon}}}{\sqrt{\pi  \hspace{0.1cm}\epsilon} } \hspace{0.1cm} r_{s} + r_{s}  \hspace{0.1cm}(r_{s} - 6  \hspace{0.1cm}m) \Big]> 0 .\nonumber \\
\label{stability11}
\end{eqnarray}
In the limit $\frac{r}{\sqrt{\epsilon}} \rightarrow \infty$ , the above Eqn.(\ref{stability11}) reduces to the following form,
\begin{equation}
\frac{m \hspace{0.1cm}(r_s - 6\hspace{0.1cm} m)}{r_s \hspace{0.1cm}(r_s - 2 \hspace{0.1cm}m)} > 0,
\end{equation}
which in turn implies that $r_s > 6 \hspace{0.1cm}m$.
Further, using the explicit form of $E$ and $L$ from Eqns. (\ref{energy1}) and (\ref{angular1}), the components of four velocity reads as below,
\begin{eqnarray}
u^{t} = \sqrt{\frac{- 2}{r A(r_s) - 2 f(r_s)}},\nonumber \\
u^{\phi} = \sqrt{\frac{- A(r_s)}{r_s \left[r_s A(r_s) - 2 f(r_s)\right]}}.
\label{first intgrl_1}
\end{eqnarray}
Using these two equations, the angular velocity of particles in these circular paths comes out as,
\begin{equation}
\Omega = \sqrt{\frac{A(r_s)}{2 r_s}}.
\label{angular velocity}
\end{equation}
Since $D_{+} = - D_{-}$, the redshift $z_{1} = z_{red}$ and blueshift $z_{2} = z_{blue}$ are equal but with  opposite signatures, the general expression for $z_1$ reads,
\begin{equation}
z_{1} = \frac{- u^{t}_e \hspace{0.1cm} u^{\phi}_d \hspace{0.1cm} D_{d_{+}} + u^{t}_d \hspace{0.1cm} u^{\phi}_e  \hspace{0.1cm} D_{e_{+}}}{u^{t}_d\hspace{0.1cm} (u^{t}_d + u^{\phi}_d \hspace{0.1cm}D_{d_{+}})}.
\label{redshift}
\end{equation}
Further, if the detector is located far away from the central object $r_d \rightarrow \infty$, Eq. (\ref{redshift}) becomes,
\begin{equation}
z_{red} =  u^{\phi}_e D_{e_{+}} = \left[\frac{ - r_{s} \hspace{0.1cm} A(r_{s})}{f(r_{s}) \Big(r_{s} \hspace{0.1cm} A(r_{s}) - 2 \hspace{0.1cm} f(r_{s}) \Big)}\right]^{\frac{1}{2}}.
\label{redshift_gen}
\end{equation}
The Eqn.(\ref{redshift_gen}) for NCBH spacetime however reduces as given below,
\begin{equation}
z_{red} = \left[-\frac{\frac{2\hspace{0.1cm} e^{\frac{-r_{s}^2}{4\hspace{0.1cm}\epsilon}\hspace{0.1cm} m\hspace{0.1cm} r_{s}^2}}{\pi \hspace{0.1cm}\epsilon^{3/2}} + \frac{4\hspace{0.1cm} m}{r_{s} \hspace{0.1cm}\sqrt{\pi}}\hspace{0.1cm}\gamma\left(\frac{3}{2}, \frac{r_{s}^2}{4 \hspace{0.1cm}\epsilon}\right)}{\left(1-\frac{4\hspace{0.1cm} m}{r_{s} \sqrt{\pi}}\gamma\left(\frac{3}{2}, \frac{r_{s}^2}{4\hspace{0.1cm} \epsilon}\right) \right)\left(\frac{2 \hspace{0.1cm}e^{\frac{-r_{s}^2}{4\hspace{0.1cm}\epsilon} \hspace{0.1cm}m \hspace{0.1cm}r_{s}^2}}{\pi \hspace{0.1cm}\epsilon^{3/2}} + \frac{12 \hspace{0.1cm}m}{r_{s} \hspace{0.1cm}\sqrt{\pi}}\hspace{0.1cm}\gamma\left(\frac{3}{2}, \frac{r_{s}^2}{4\hspace{0.1cm} \epsilon}\right)-2\right)}\right]^{\frac{1}{2}}.
\label{redshift_mass}
\end{equation}
The bound on $r_s$ and $z$ with $\epsilon$ is shown in Table.\ref{table1}. One can observe from the Table.\ref{table1} that with increasing value of non-commutative parameter, the radius of stable circular orbits as well as frequency shift increases, which implies that beyond a particular value of radius, the orbits will always be stable and the emitted photons by massive particles following these stable circular orbits will always be red shifted.
\begin{table}[ht]
\caption{Bound on radius and frequency shift for different values of $\epsilon$ for $m = 1$. }
\centering
\begin{tabular}{| c | c | c | c | c |}
\hline 
Non-commutative parameter ($\epsilon$) & \hspace{1.6cm} 1 &\hspace{1.6cm} 3 & \hspace{1.6cm}9 & \hspace{1.6cm}15 \\ [1ex]
\hline 
Radius of stable circular orbit ($r_s$) &\hspace{1.6cm} 6.20 & \hspace{1.6cm} 10.46 & \hspace{1.6cm}16.94 & \hspace{1.6cm}20.52\\[1ex]
\hline 
Frequency shift ($z$)  & \hspace{1.6cm} 0.017 & \hspace{1.6cm} 0.025 & \hspace{1.6cm} 0.035 & \hspace{1.6cm} 0.046 \\ [1ex]
\hline 
\end{tabular} 
\label{table1}
\end{table}
In limit $\frac{r}{\sqrt{\epsilon}} \rightarrow \infty$, the relation (\ref{redshift_mass}) reduces into the standard SBH case \cite{bec}. 
\subsection{Mass parameter of NCBH from red/blue shift}
\begin{figure}
\centerline{
\includegraphics[width=9cm,height=6cm]{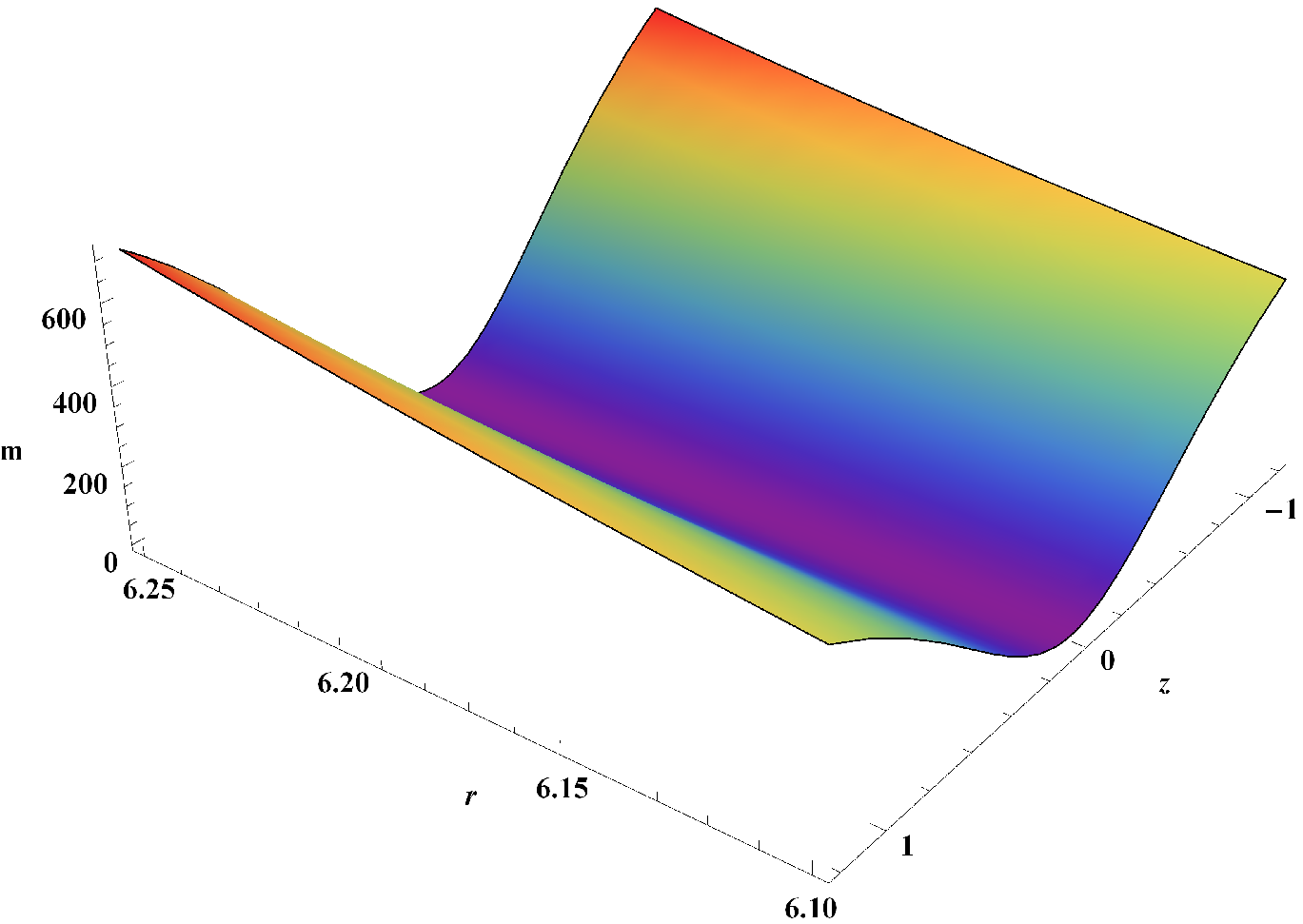}
\includegraphics[width=9cm,height=6cm]{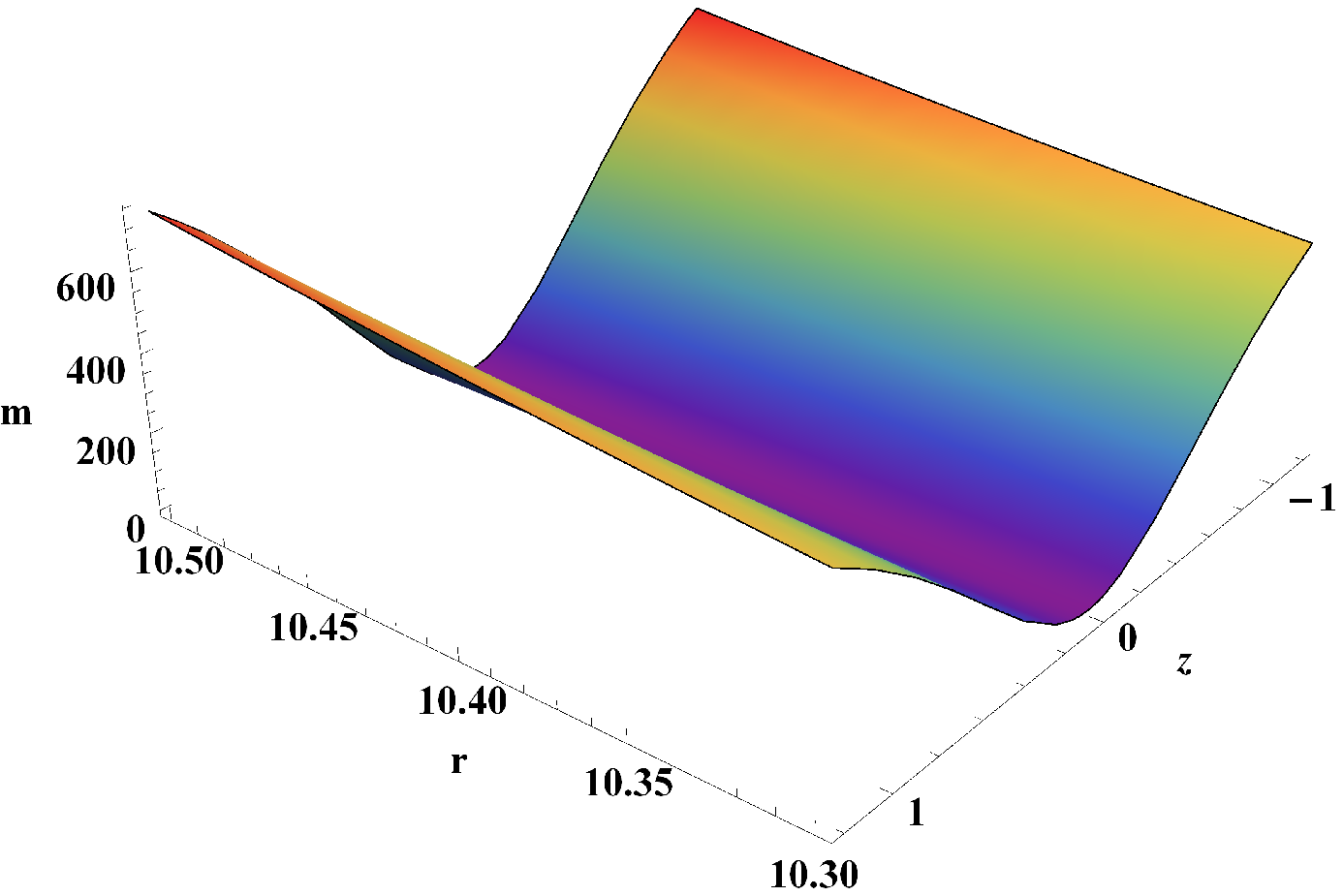}}
\caption{Variation of mass parameter ($m$), redshift ($z$) and radius of circular orbit ($r$) for $\epsilon = 1$ (left panel) and $\epsilon = 3$ (right panel)}
\label{lastfig}
\end{figure}

\begin{figure}
\centerline{
\includegraphics[width=9cm,height=6cm]{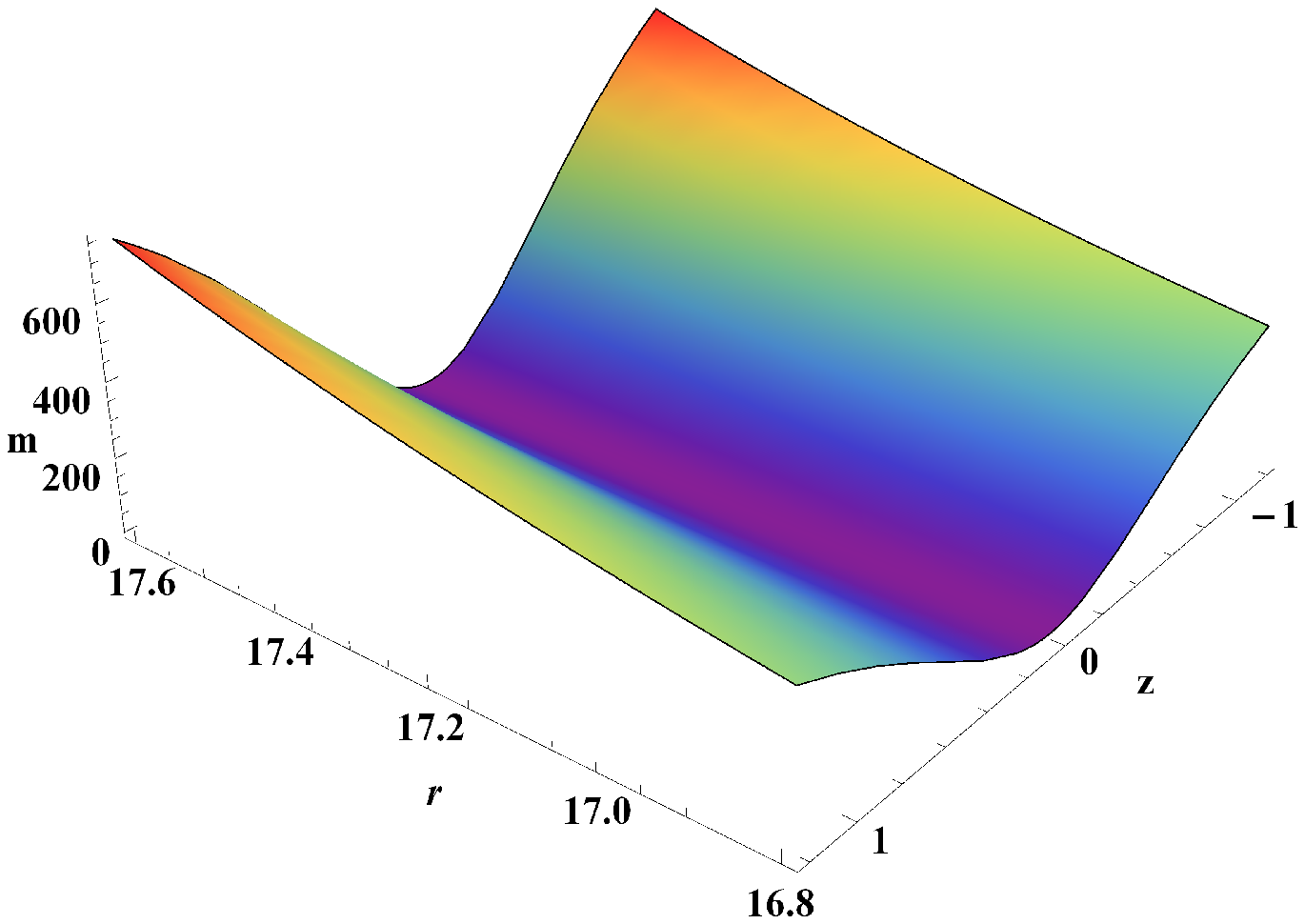}
\includegraphics[width=9cm,height=6cm]{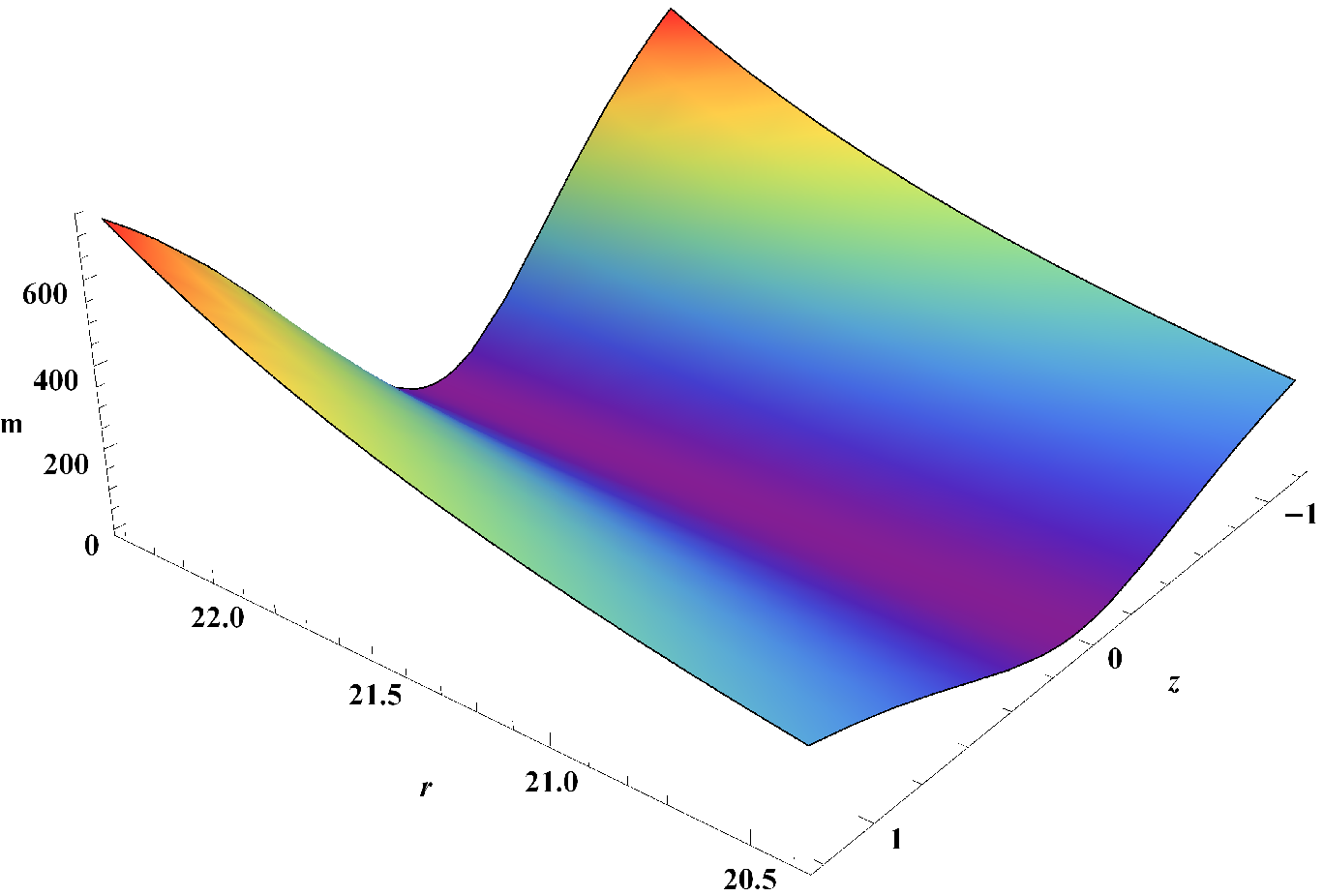}}
\caption{Variation of mass parameter ($m$), redshift ($z$) and radius of circular orbit ($r$) for $\epsilon = 9$ (left panel) and $\epsilon = 15$ (right panel)}
\label{lastfig1}
\end{figure}

\noindent Here, we have expressed the mass parameter of a NCBH in terms of the redshift of photons emitted by a particle moving along the circular geodesics as given below,
\begin{eqnarray}
m = \frac{1}{8 \hspace{0.1cm}z^2\hspace{0.1cm} \gamma\left(\frac{3}{2}, \frac{r^2}{4 \hspace{0.1cm}\epsilon}\right) \Big(6 \hspace{0.1cm}\gamma\left(\frac{3}{2}, \frac{r^2}{4 \hspace{0.1cm}\epsilon}\right) \sqrt{\pi}\hspace{0.1cm} \epsilon^{3/2}  + e^{\frac{-r^2}{4 \hspace{0.1cm}\epsilon}} \hspace{0.1cm}r^3\Big)} \times \nonumber \\
\Bigg[ (1 + z^2) \hspace{0.1cm}(e^{\frac{-r^2}{4 \hspace{0.1cm}\epsilon}} \sqrt{\pi}\hspace{0.1cm} r^3) + 
 10 \hspace{0.1cm}z^2 \hspace{0.1cm}\pi \hspace{0.1cm}\epsilon^{3/2}\hspace{0.1cm} \gamma\left(\frac{3}{2}, \frac{r^2}{4 \hspace{0.1cm}\epsilon}\right) + 2\hspace{0.1cm} \pi \hspace{0.1cm}\gamma\left(\frac{3}{2}, \frac{r^2}{4\hspace{0.1cm} \epsilon}\right) \epsilon^{3/2} \nonumber \\
- \sqrt{\pi}\hspace{0.1cm} \Big( (1 + 2 z^2 + z^4)\hspace{0.1cm} r^6\hspace{0.1cm} e^{\frac{-r^2}{2 \hspace{0.1cm}\epsilon}} + (1 + 6 \hspace{0.1cm}z^2 + z^4) \hspace{0.1cm}4 r^3 \sqrt{\pi} \hspace{0.1cm}\epsilon^{3/2} \hspace{0.1cm}e^{\frac{-r^2}{4 \hspace{0.1cm}\epsilon}}\gamma\left(\frac{3}{2}, \frac{r^2}{4 \hspace{0.1cm}\epsilon}\right) + \nonumber \\
(1 + 10 z^2 +  z^4)\hspace{0.1cm} 4 \hspace{0.1cm}\pi \hspace{0.1cm}\left(\gamma\left(\frac{3}{2}, \frac{r^2}{4 \hspace{0.1cm}\epsilon}\right)\right)^2 \hspace{0.1cm}\epsilon^{3}\Big)^{1/2}\Bigg] r.
\label{redshift_mass_1}
\end{eqnarray}
The above Eqn.(\ref{redshift_mass_1}) is a relation between the red-shift, the mass parameter of a NCBH, the non-commutative parameter $\epsilon$ and the radius of circular orbit. Hence, a measurement of the redshift of light emitted by a particle that follows a circular orbit in equatorial plane around a NCBH will have a mass parameter determined by Eqn.(\ref{redshift_mass_1}). In Fig.(\ref{lastfig}) and Fig.(\ref{lastfig1}), the mass parameter $m$ as a function of redshift and the radius of circular orbit of a photon is shown. Here, the bound on radius of stable circular orbit is calculated from Eqn.(\ref{stability11}). The variation of mass parameter with redshift and radius is qualitatively same as for the SBH spacetime whereas on increasing the value of non-commutative parameter the scale of mass parameter decreases. 
One can observe that in the limit $\frac{r}{\sqrt{\epsilon}} \rightarrow \infty$,
the relation (\ref{redshift_mass_1}) reduces into the standard SBH case \cite{bec}.
\section{Summary and Conclusions}
\noindent In this article, we have investigated the geodesic motion of massless particles i.e. photons and frequency shift of photons in the background of a NCBH spacetime. Some of the interesting results obtained are summarised below.
\begin{enumerate}[label=(\roman*)]
\item The effective potential of the non-commutative BH  near the origin ($r\rightarrow0$) differs from the Schwarzschild case because of the noncommutativity. One can observe from the respective plot of potential that, with the increase in the value of non-commutative parameter circular orbit shifts away from the central object.
\item It is observed that the massless test particles can be trapped by the NCBH. The various types of orbits (i.e. unstable circular orbit, fly-by orbit, terminating escape orbit and terminating bound orbit) are present for the massless particles, corresponding to the particular value of the impact parameter. 
\item It is found that the radius of unstable circular orbit for NCBH is smaller than SBH i.e the gravitational field of a NCBH spacetime is weaker than that of SBH spacetime.
\item It is also observed that the time periods for unstable circular orbit is smaller for NCBH as compared to the SBH which is consistent with the fact that the non-commutative parameter is responsible to decrease the gravitational strength of a NCBH case.
\item  The nature of frequency shift of photons emitted by particles moving along circular geodesics is qualitatively similar for different values of the non-commutative parameter. However, the gravitational field of a NCBH is found to be less attractive in nature in comparison to SBH.
\item The effect of non-commutative parameter is very small but it is important since it manifests the nature of spacetime structure at quantum gravity level. We conclude that the gravitation field at quantum level is weaker than that in classical scenario.
\end{enumerate}
In future, we plan to investigate the null geodesic structure of the non-commutative inspired charged and rotating BH spacetimes.
\acknowledgements 
\noindent The authors (RSK and HN) would like to thank Department of Science and Technology, New Delhi for the financial support through grant no. SR/FTP/PS-31/2009. One of the author RU is also thankful to IUCAA, Pune for support under its visitor program where a part of this work was carried out.


\begin{references}
\bibitem{har} J. B. Hartle.: \textit{Gravity: An Introduction To Einstein's General Relativity}. (Pearson Education Inc., Singapore, 2003).
\bibitem{wal} R. M. Wald.: \textit{General Relativity}. (University of Chicago Press, Chicago, USA, 1984).
\bibitem{psj} P. S. Joshi.: \textit{Global aspects in gravitation and cosmology}. (Oxford University Press, Oxford, UK, 1997).
\bibitem{cha} S. Chandrasekhar.: \textit{The Mathematical Theory of Black Holes}. (Oxford University Press, New York, 1983).
\bibitem{epo} E. Poisson.: \textit{A Relativist's Toolkit: The Mathematics of Black-Hole Mechanics}. (Cambridge University Press, Cambridge, 2004).
\bibitem{nic} P. Nicolini,  A. Smailagic and E. Spallucci.: \textit{Noncommutative geometry inspired Schwarzschild black hole}. Phys. Lett. \textbf{ B632}, 547-551 (2006).
\bibitem{rah} F. Rahman, I. Radinschi, U. F. Mondal and P. Bhar.: \textit{Particle's Motion Around a Non-Commutative Black Hole}. Int. J. Theor. Phys. \textbf{54}, 1038-1051 (2015).
\bibitem{piy} P. Bhar, F. Rahman, R. Biswas and U. F. Mondal.: \textit{Particles and scalar waves in noncommutative charged black hole spacetime}. Commun. Theor. Phys. \textbf{64}, 1-8 (2015).
\bibitem {RefP1} M. P. Dabrowski and A. L. Larsen: \textit{Null strings in Schwarzschild space-time}. Phys. Rev. \textbf{D55}, 6409-6414 (1997).
\bibitem {RefF1} S. Fernando.: \textit{Null geodesics of Charged Black Holes in String Theory}. Phys. Rev. \textbf{D85},  024033  (2012).
\bibitem {RefN} N. Cruz, M. Olivares and R. J. Villanueva.: \textit{The geodesic structure of the Schwarzschild Anti-de Sitter black hole}. Class. Quant. Grav. \textbf{22}, 1167-1190 (2005).
\bibitem {RefD1} D. Pugliese, H. Quevedo and R. Ruffini.: \textit{Circular motion of neutral test particles in Reissner-Nordstr$\ddot{o}$m spacetime}. Phys. Rev. \textbf{D43}, 3140 (1991).
\bibitem {RefF2} S. Fernando.: \textit{Schwarzschild black hole surrounded by quintessence: Null geodesics}. Gen. Rel. Grav. \textbf{44}, 1857 (2012).
\bibitem {RefF3} S. Fernando, S. Meadows and K. Reis.: \textit{Null trajectories and bending of light in charged black holes with quintessence}. \textbf{arXiv:1411.3192 [gr-qc]}  (2014).
\bibitem {RefK1} K. Hioki and U. Miyamoto.: \textit{Hidden symmetries, null geodesics, and photon capture in the Sen black hole}. Phys. Rev. \textbf{D78}, 044007 (2008).
\bibitem {RefB} A. Bhadra.: \textit{Gravitational lensing by a charged black hole of string theory}. Phys. Rev. \textbf{D67}, 103009 (2003).
\bibitem {RefP2} P. Pradhan and P. Majumdar.: \textit{Circular Orbits in Extremal Reissner Nordstr$\ddot{o}$m Spacetimes}. Phys. Lett. \textbf{A375}, 474-479 (2011).
\bibitem {RefP3} P. Pradhan.: \textit{Circular  Geodesics in the Kerr-Newman-Taub-NUT Space-time}. Class. Quantum Grav. \textbf{32}, 165001 (2015).
\bibitem {RefZ} Z. Stuchl$\acute{i}$k and  J. Schee.: \textit{Circular geodesic of Bardeen and Ayon-Beato-Garcia regular black-hole and no-horizon spacetimes}. Int. J. Mod. Phys. \textbf{D24}, 1550020 (2014).
\bibitem {RefJ} J. Schee and Z. Stuchlik.: \textit{Profiles of emission lines generated by rings orbiting braneworld Kerr black holes}. Gen. Rel. Grav. \textbf{41}, 1795-1818 (2009).
\bibitem {RefK2} R. Koley, S. Pal and S. Kar.: \textit{Geodesics and geodesic deviation in a two-dimensional black hole}. Am. J. Phys. \textbf{71}, 1037 (2003).
\bibitem {RefU} R. Uniyal, H. Nandan and K. D. Purohit.: \textit{Geodesic Motion in a Charged $2$D Stringy Black Hole Spacetime}. Mod. Phys. Lett. \textbf{A29}, 1450157 (2014).
\bibitem {RefU1} R. Uniyal, N. C. Devi, H. Nandan and K. D. Purohit.: \textit{Geodesic Motion in Schwarzschild Spacetime Surrounded by Quintessence}. Gen. Rel. Grav. \textbf{47}, 16 (2015).
\bibitem {RefE1} E. Hackmann.: \textit{Geodesic equations in black hole space-times with cosmological constant}. Ph. D. Thesis, (University of Bremen, Germany) (2010).
\bibitem {RefT1} T. Maki and K. Shiraishi.: \textit{Motion of test particles around a charged dilatonic black hole}. Class. Quant. Grav. \textbf{11}, 227-238 (1994).
\bibitem {RefU2} R. Uniyal, A. Biswas, H. Nandan and K. D. Purohit.: \textit{Geodesic motion in R-charged black hole spacetimes}. Phys. Rev. \textbf{D92}, 8, 084023 (2015).
\bibitem {RefRS} R. S. Kuniyal, R. Uniyal, H. Nandan and K. D. Purohit.: \textit{Null Geodesics in a Magnetically Charged Stringy Black Hole Spacetime}. Gen. Rel. Grav. \textbf{48}, 46 (2016).
\bibitem {RefF4} S. Fernando, D. Krug and  C. Curry.: \textit{Geodesic structure of static charged black hole solutions in 2+1 dimensions}. Gen. Rel. Grav. \textbf{35}, 1243-1261 (2003).
\bibitem {RefE2} E. Hackmann, V. Kagramanova, J. Kunz and C. Lammerzahl.: \textit{Analytic solutions of the geodesic equation in higher dimensional static spherically symmetric space-times}. Phys. Rev. \textbf{D78}, 124018 (2008).
\bibitem {RefH1} A. Dasgupta, H. Nandan and S. Kar.: \textit{Kinematics of deformable media}. Annals Phys. \textbf{323}, 1621-1643 (2008).
\bibitem {RefA1} A. Dasgupta, H. Nandan, and S. Kar.: \textit{Kinematics of flows on curved, deformable media}. Int. J. Geom. Meth. Mod. Phys. \textbf{6}, 645-666 (2009).
\bibitem {RefA3} A. Dasgupta, H. Nandan and S. Kar.: \textit{Geodesic flows in rotating Black hole backgrounds}. Phys. Rev. \textbf{D85}, 104037 (2012).
\bibitem {RefG1} S. Ghosh, S. Kar and H. Nandan.: \textit{Confinement of test particles in warped spacetimes}. Phys. Rev. \textbf{D82}, 024040 (2010).
\bibitem {RefR} R. S. Kuniyal, R. Uniyal, H. Nandan and A. Zaidi.: \textit{Geodesic flows around charged black holes in two dimensions}. Astrophys. Space Sci. \textbf{357}, 92 (2015).
\bibitem {RefR11} R. Uniyal and H. Nandan.: \textit{Geodesic flows in a Charged Black Hole Spacetime with Quintessence}. \textbf{arXiv:1612.07455 v1 [gr-qc]} (2016).
\bibitem{ras22} R. Uniyal.: \textit{Geodesic congruences around various spacetime backgrounds}. Ph. D. Thesis, (Gurukula Kangri Vishwavidyalaya, Haridwar, Uttarakhand, India) (2016).
\bibitem{ehm1} E. Hackmann and Hongxiao Xu.: \textit{Charged particle motion in Kerr-Newmann space-times}. Phys. Rev. \textbf{D87},124030 (2013).
\bibitem{ehm} E. Hackmann, C. L$\ddot{a}$mmerzahl, Y. N. Obukhov, D. Puetzfeld and I. Schaffer.: \textit{Motion of spinning test bodies in Kerr spacetime}. Phys. Rev. \textbf{D90}, 064035 (2014).
\bibitem{dab} M. P. Dabrowski and F. E. Schunck.: \textit{Boson Stars as Gravitational Lenses}. Astrophysics. J. \textbf{535}, 316-324 (2000).
\bibitem{hnd} H. Nandan, N. M. Bezares-Roder and H. Dehnen.: \textit{Black Hole Solutions and Pressure Terms in Induced Gravity with Higgs Potential}. Class. Quant. Grav. \textbf{27}, 245003 (2010).
\bibitem{val} V. Diemer, K. Eilers, B. Hartmann, I. Schaffer and C. Toma.: \textit{Geodesic motion in the space-time of a non-compact boson star}. Phys. Rev. \textbf{D88},  044025 (2013).
\bibitem{sas1} S. Grunau and B. Khamesra.: \textit{Geodesic motion in the (rotating) black string spacetime}. Phys. Rev. \textbf{D87}, 124019 (2013).
\bibitem{sas2} S. Grunau and V. Kagramanova.: \textit{Geodesics of electrically and magnetically charged test particles in the Reissner-Nordstr$\ddot{o}$m space-time: analytical solutions }. Phys. Rev. \textbf{D83}, 044009 (2011).
\bibitem{rpk} R. S. Kuniyal, H. Nandan, U. Papnoi, R. Uniyal and K. D. Purohit.: \textit{Strong gravitational lensing in 5D Myers-Perry black hole spacetime} (\textbf{Submitted for publication}).
\bibitem{raji} R. Shaikh, S. Kar and A. Dasgupta.: \textit{Evolution of geodesic congruences in a gravitationally collapsing scalar field background}. Phys. Rev. \textbf{D90}, 124069 (2014).
\bibitem{kaif} K. Flathmann and S. Grunau.: \textit{Analytic solutions of the geodesic equation for Einstein-Maxwell-dilaton-axion black holes}. Phys. Rev. \textbf{D92}, 104027 (2015).
\bibitem{sahe} S. Soroushfar, R. Saffari, S. Kazempour, S. Grunau and J. Kunz.: \textit{Detailed study of geodesics in the Kerr-Newman-(A)ds spacetime and the rotating charged black hole spacetime in $f(R)$ gravity}. Phys. Rev. \textbf{D94}, 024052 (2016).
\bibitem{bro} E. Brown and R. B. Mann.: \textit{Instability of the Noncommutative Geometry Inspired Black Hole}. Phys. Lett. \textbf{B694}, 440-445 (2011).
\bibitem{arr} I. Arraut, D. Batic and M. Nowakowski.: \textit{A Non commutative model for a mini black hole}. Class. Quant. Grav. \textbf{26}, 245006 (2009).
\bibitem{kli} C. Klimicik, P. Konik and A. Pompos.: \textit{Non-commutative black holes in D dimensions}. \textbf{arXiv:9405012 v1 [gr-qc]} (1994).
\bibitem{sma} A. Smailagic and E. Spallucci.: \textit{Lorentz invariance, unitarity in UV-finite of QFT on noncommutative spacetime}. J. Phys. \textbf{A37}, 1-10 (2004).
\bibitem{gre} E. Di Grezia, G. Esposito and G. Miele.: \textit{Black hole evaporation in a spherically symmetric non-commutative space-time}. J. Phys. \textbf{A41}, 164063 (2008).
\bibitem{anso} S. Ansoldi, P. Nicolini, A. Smailagic and E. Spallucci.: \textit{Noncommutative geometry inspired charged black holes}. Phys. Lett. \textbf{B645}, 261-266 (2007).
\bibitem{alav} S. A. Alavi.: \textit{Reissner-Nordstr$\ddot{o}$m black hole in noncommutative spaces}. Acta Phys. Polon. \textbf{B40}, 2679-2687 (2009).
\bibitem{grezi} E. Di Grezia and G. Esposito.: \textit{Non-commutative Kerr black hole}. Int. J. Geom. Meth. Mod. Phys. \textbf{08}, 657-668 (2011).
\bibitem{modesto} L. Modesto and P. Nicolini.: \textit{Charged rotating noncommutative black holes}. Phys. Rev. \textbf{D82}, 104035 (2010).
\bibitem{chabab} M. Chabab, H. EI Moummi and M. B. Sedra.: \textit{On Schwarzschild black holes in a D-dimensional noncommutative space}. \textbf{arXiv:1201.2547 v1 [hep-th]} (2012).
\bibitem{rah1} F. Rahaman, P. K. F. Kuhfitting, B. C. Bhui. M. Rahaman, S. Ray and U. F. Mondal.: \textit{BTZ black holes inspired by noncommutative geometry}. Phys. Rev. \textbf{D87}, 084014 (2013).
\bibitem{alf15} A. Herrera-Aguilar and U. Nucamendi.: \textit{Kerr black hole parameters in terms of red/blue shifts of photons emitted by geodesic particles}. \textbf{arXiv:1506.05182 v1 [gr-qc]} (2015).
\bibitem{bec} R. Becerril, S. Valdez-Alvarado and U. Nucamendi.: \textit{Obtaining mass parameter of compact objects from red-blue shifts emitted by geodesic particles around them}. Phys. Rev. \textbf{D94}, 124024 (2016).
\end{references}
\end{document}